\documentclass[submission, Phys]{SciPost}
\hypersetup{
    colorlinks,
    linkcolor={red!50!black},
    citecolor={blue!50!black},
    urlcolor={blue!80!black}
}

\usepackage{makecell}
\usepackage[bitstream-charter]{mathdesign}
\usepackage{physics}
\usepackage{dsfont}
\usepackage{graphicx}
\usepackage{comment}
\usepackage{caption}
\usepackage{subcaption}
\usepackage{multirow}
\usepackage{booktabs}
\graphicspath{{Figures/}}
\newcommand{\ie}{\hbox{\em i.e.{}}}
\newcommand{\bcdot}{\boldsymbol{\cdot}}
\renewcommand{\tr}{\mathrm{Tr}}
\DeclareSymbolFont{usualmathcal}{OMS}{cmsy}{m}{n}
\DeclareSymbolFontAlphabet{\mathcal}{usualmathcal}

\fancypagestyle{SPstyle}{
\fancyhf{}
\lhead{\colorbox{scipostblue}{\bf \color{white} ~SciPost Physics}}
\rhead{{\bf \color{scipostdeepblue} ~Submission }}

\fancyfoot[C]{\textbf{\thepage}}
}

\begin{document}

\pagestyle{SPstyle}

\begin{center}{\Large \textbf{\color{scipostdeepblue}{
Coherent Generation and Protection of Anticoherent Spin States 
}}}\end{center}

\begin{center}\textbf{
Jérôme Denis\textsuperscript{$\star$},
Colin Read\textsuperscript{$\dagger$} and
John Martin\textsuperscript{$\ddagger$}
}\end{center}

\begin{center}
Institut de Physique Nucléaire, Atomique et de Spectroscopie, CESAM, University of Liège, B-4000 Liège, Belgium\\[\baselineskip]
$\star$ \href{mailto:email1}{\small jdenis@uliege.be}\,,\quad
$\dagger$ \href{mailto:email2}{\small cread@uliege.be}\,,\quad
$\ddagger$ \href{mailto:email2}{\small jmartin@uliege.be}
\end{center}

\section*{\color{scipostdeepblue}{Abstract}}
\boldmath\textbf{%
We report the first protocol specifically designed to generate anticoherent spin-$j$ states at different orders. The protocol consists of cycles of a rotation pulse about an axis followed by a squeezing pulse in a perpendicular direction. To protect these states, we develop dynamical decoupling techniques using group-based sequence design and the dynamically corrected gate formalism. We analyze key sources of dephasing, disorder, and dipole-dipole interactions and assess the effectiveness of our methods in preserving coherence. Potential applications of the produced anticoherent spin states include quantum sensing and studies related to quantum entanglement.
}

\vspace{\baselineskip}

\noindent\textcolor{white!90!black}{%
\fbox{\parbox{0.975\linewidth}{%
\textcolor{white!40!black}{\begin{tabular}{lr}%
  \begin{minipage}{0.6\textwidth}%
    {\small Copyright attribution to authors. \newline
    This work is a submission to SciPost Physics. \newline
    License information to appear upon publication. \newline
    Publication information to appear upon publication.}
  \end{minipage} & \begin{minipage}{0.4\textwidth}
    {\small Received Date \newline Accepted Date \newline Published Date}%
  \end{minipage}
\end{tabular}}
}}
}


\vspace{10pt}
\noindent\rule{\textwidth}{1pt}
\tableofcontents
\noindent\rule{\textwidth}{1pt}
\vspace{10pt}

\section{Introduction}
Anticoherent (AC) spin states represent an extreme form of quantum behavior, contrasting sharply with coherent spin states, which closely resemble classical angular momentum states (see \cite{Zim:06} for the origin of the concept and Section \ref{sec:ACstates} for a detailed introduction). While coherent spin states exhibit minimal quantum fluctuations, pure AC states maximize quantum uncertainty, making them fundamentally distinct. This exacerbated quantum nature makes AC states highly sensitive to external perturbations, a feature that, while advantageous for certain applications~\cite{ZGoldberg2021}, also makes them particularly fragile in the presence of decoherence~\cite{Denis_2022}. 

The interest in AC states stems from their unique property of achieving equal sensitivity to rotations around any axis~\cite{2017Chryssomalakos,2018Goldberg,2020Martin,2025ESE,2025Goldberg}. This rotational invariance not only makes them ideal for tasks such as the alignment of Cartesian reference frames~\cite{Kolenderski2008}, but also positions them as powerful tools in the wider field of quantum metrology. In particular, non-classical states of atomic ensembles, such as squeezed spin states and highly entangled states, are known to beat the standard quantum limit in terms of sensitivity~\cite{Pezz2018}. However, while squeezed states generally improve accuracy along one axis at the expense of others, AC states, of which the spin-$2$ tetrahedron state is an example, uniquely enable the simultaneous estimation of multiple parameters~\cite{2024Ferretti}. This capability stems from their high degree of symmetry, often associated with Platonic solids, which makes them particularly advantageous when the direction of the applied transformation is unknown~\cite{2018Goldberg,2020Martin,2025ESE}. Ultimately, by allowing simultaneous estimation of several transformation parameters with precision at the Heisenberg limit, AC states offer a significant advantage in quantum-enhanced measurement strategies.

Schrödinger cat states and their multipartite equivalent, the Greenberger-Horne-Zeilinger (GHZ) states, are well-known examples of AC states, though they only exhibit first-order anticoherence. They have been successfully created on various physical platforms, such as photons, neutral atoms, and spins, and for different numbers of constituents or spin values; see, for example,~\cite{Pan2000,Omran2019,Pont2024,Yu2025,2023Bhatti,2019Zhu} and references therein. However, higher-order AC states have so far been generated exclusively in multiphotonic systems~\cite{2017Bouchard,2024Ferretti}, underscoring the urgent need for protocols that enable their realization on other physical platforms, such as atomic or solid-state systems. In this work, we fill this gap by introducing simple protocols that enable the generation of AC states of various orders using only spin rotation and squeezing.

This manuscript is organized as follows. In Section \ref{sec:ACstates}, we review the concept of AC spin states and discuss their general properties and interest. Section \ref{sec:protocolACStates} details our protocol for generating AC states of different orders. Section \ref{Sec.Dec.} examines decoherence and mitigation strategies, covering typical sources of dephasing and dynamical decoupling techniques. Section \ref{sec:RobustACStates} discusses robust AC state generation, including the design of dynamically corrected gates, error-resistant protocols using finite-duration pulses, and the impact of control errors. Finally, Section \ref{sec:Conclusion} summarizes our conclusions and outlines possible directions for future work. Additional technical details are provided in the appendices.

\section{Anticoherent spin states}
\label{sec:ACstates}
In a general study of individual spin-$j$ systems, anticoherence is best defined based on the density operator, as it offers a complete description of the system’s state, including both pure and mixed cases. 
The density operator $\rho$ can be expressed in terms of multipolar tensor operators, defined as
\begin{equation}
    T_{LM}=\frac{\sqrt{(2j-L)!(2j+L+1)!}}{\sqrt{4\pi}\,(2j)!}\int_{\mathcal{S}^2}Y_{LM}(\Omega)\,|\Omega\rangle\langle\Omega|\,d\Omega,
\end{equation}
where $L = 0,1,\dots,2j$ and $M = -L,\dots,L$. Here, $|\Omega\rangle$ denotes a spin-coherent state oriented along the direction $\Omega \equiv (\theta,\varphi)$~\cite{Zare}. The multipolar tensor operators are in direct correspondence with the spherical harmonics $Y_{LM}(\Omega)$, acting as their operator analogues. They transform according to the $(2L+1)$-dimensional irreducible representations of the spin rotation group $\mathrm{SU}(2)$ and form a complete orthonormal set under the Hilbert-Schmidt inner product. The expansion of $\rho$ then takes the form
\begin{equation}\label{rhoexpansion}
    \rho = \sum_{L=0}^{2j} \sum_{M=-L}^{L} \rho_{LM} T_{LM},
\end{equation}
where the coefficients $\rho_{LM}$ are known as multipole moments or statistical tensors~\cite{Var.Mos.Khe:88,Biedenharn1984,1952Schwinger,Zare}. The multipoles $\rho_{LM}$ are measurable physical quantities that encode information about the system's polarization and coherence and can be determined from intensity moments~\cite{Goldberg2022}. The tensor operators basis thus serves as a powerful tool for both theoretical and experimental investigations of spin-$j$ systems. The quantum number $L$ of the tensor operator corresponds to the multipole order, with $L = 0$ representing the monopole (population), $L = 1$ the dipole (orientation), $L = 2$ the quadrupole (alignment), etc., for higher orders. The monopole component corresponds to a scalar quantity that remains invariant under rotations. Because it has no angular dependence, it is the simplest and most isotropic term. The dipole component is a vector-type quantity that transforms under rotations like a vector in a three-dimensional space. A nonzero dipole moment indicates a preferred axis along which the spin expectation value is aligned. The quadrupole component is a rank-$2$ tensor that characterizes the shape of the angular distribution rather than its net direction. A nonzero quadrupole moment means that the state exhibits alignment rather than mere orientation, i.e., the spin projections favor particular axes without necessarily having a net spin vector. Based on this formalism, we can now define anticoherent spin states.

\paragraph{Definition} A state $\rho$ is said to be anticoherent to order $t$, or $t$-AC, if $\rho_{LM}=0\:\forall\,M$ for $1\leq L\leq t$, which means that all multipole moments of order $L$ up to $t$ vanish.\\

Let $\mathbf{J}=(J_x,J_y,J_z)$ denote the spin operator. An AC state to order $1$ is defined by the absence of net orientation, i.e.\ $\langle \mathbf{J}\rangle=0$, while an AC state to order $2$ further requires isotropic fluctuations, $\Delta J_x^2=\Delta J_y^2=\Delta J_z^2$. More generally, an AC state to order $t$ is one for which all multipole moments up to rank $L=t$ vanish, except for the monopole term fixed by normalization. This leads to a more isotropic statistical distribution of the angular momentum components and their products and makes AC states particularly suitable for applications that require rotational invariance. A remarkable feature of pure spin states is their ability to exhibit anticoherence up to an order limited by the spin quantum number $j$. Such AC states, referred to as "kings of quantumness" in Ref.~\cite{2015SanchezSoto}, contain no lower-order multipolar contributions in the expansion \eqref{rhoexpansion}.  To illustrate this, consider two examples using the common eigenbasis of $J^2$ and $J_z$, formed by the $|j, m\rangle$ states, which satisfy $J^2 |j, m\rangle = j(j+1) |j, m\rangle$ and $J_z |j, m\rangle = m |j, m\rangle$ (we set $\hbar=1$). First, the spin cat state
\begin{equation}\label{SCS}
|\psi_{\mathrm{cat}}\rangle=\frac{1}{\sqrt{2}}(|j,j\rangle+|j,-j\rangle)
\end{equation}
is anticoherent to order $1$ for any $j>1/2$, since $\langle \mathbf{J}\rangle=0$, but it does not exhibit higher-order anticoherence.
By contrast, the spin-$2$ tetrahedron state
\begin{equation}\label{spin2tetra}
|\psi_{\mathrm{tetra}}\rangle=\frac{1}{2}(|2,2\rangle+i\sqrt{2}|2,0\rangle+|2,-2\rangle)
\end{equation}
is anticoherent to order $2$, satisfying both $\langle \mathbf{J}\rangle=0$ and the isotropy condition $\Delta J_x^2=\Delta J_y^2=\Delta J_z^2$. These examples illustrate how carefully constructed superpositions can cancel specific angular momentum moments while preserving higher-order coherence, highlighting the nonclassical nature of these states.

Beyond their fundamental interest, AC states have been studied in a variety of contexts. In Majorana’s representation, a pure spin-$j$ state corresponds to a symmetric state of $N = 2j$ qubits~\cite{1932Majorana,2020ESE}. From the entanglement perspective, a spin-$j$ AC state can then be interpreted as a maximally entangled state of $N = 2j$ spin-$1/2$ particles within the symmetric subspace, shedding light on the relationship between anticoherence and multipartite entanglement~\cite{Baguette2014}. As a result, the states generated by the protocols presented in this work correspond to the most entangled symmetric multiqubit states. Their metrological usefulness has also been explored, particularly in optimizing precision for rotation estimation with both pure and mixed states. It was shown that AC states are optimal quantum rotosensors and that the higher their AC order $t$, the more advantageous their metrological advantage~\cite{2020Martin,2025Goldberg}. These properties highlight the significance of AC states both in quantum entanglement theory and in quantum metrology.


\section{Protocol for generating AC states}
\label{sec:protocolACStates}

In this section, we present our protocol for generating a sequence of control operations that produce a pure AC state of a given order $t$ in a spin-$j$ system. Our analysis focuses exclusively on the unitary dynamics driven by a time-dependent Hamiltonian, while the impact of decoherence will be addressed in the next section.

\subsection{Controls and figure of merit}

As in quantum optimal control (QOC), a crucial first step is to define an objective function that depends on the controls and serves as a figure of merit, quantifying the quality of the final state produced by them. Our goal is to prepare an anticoherent state of a given order $t$, regardless of its specific form, rather than to aim for a predetermined target state. We therefore consider measures that quantify the degree of anticoherence of spin states to a given order $t$, which we refer to as $t$-AC measures. Various such measures have been proposed; see, e.g.,~\cite{delaHoz2013,2017Baguette,Goldberg2021} and references therein, but among these, one family stands out due to its exceptional sensitivity to deviations from anticoherence: $t$-AC measures based on the Bures distance between density operators, referred to as $\mathcal{A}_t^{\mathrm{Bures}}$~\cite{2017Baguette}. Using these measures (for different orders $t$) will ensure that the pure states generated by our protocol have properties very close to those of a genuine $t$-AC state. The $t$-AC measure based on the Bures distance is defined as
\begin{equation}
    \mathcal{A}_t^{\mathrm{Bures}}(\rho)=1-\sqrt{\frac{\sqrt{t+1}-\sum_{i=1}^{t+1}\sqrt{\lambda_i}}{\sqrt{t+1}-1}}
    \label{eq:bures_acmeasure}
\end{equation}
where $\lambda_i$ are the eigenvalues of the spin-$t/2$ reduced state obtained by tracing out $2j-t$ spin-$1/2$ constituents (see~\cite{2017Baguette} for more details). In the following, we write $\mathcal{A}_t(\rho)$ for $\mathcal{A}_t^{\mathrm{Bures}}(\rho)$ to keep expressions concise. The $t$-AC measure \eqref{eq:bures_acmeasure} can take a value between $0$ and $1$, these two extreme values being realized only for coherent spin states ($\lambda_1=1$, $\lambda_{i>1}=0$) and anticoherent spin states ($\lambda_{i}=1/(t+1)$ $\forall\,i$).

To explore the controlled generation of pure AC spin states, we consider a Hamiltonian capable of producing any $\mathrm{SU}(2j+1)$ spin unitary transformation, and therefore of accessing the full state space~\cite{2003Giorda,2008Merkel}, which is of the form
\begin{equation}
H(t)=\Omega(t)\big[\cos\big(\phi(t)\big)J_{x}+\sin\big(\phi(t)\big)J_{y}\big]+\chi(t) J_{z}^{2}
    \label{eq:hamiltonian}
\end{equation}
where $\Omega(t)$ is the rotation rate about an axis in the $x$–$y$ plane oriented at an angle $\phi(t)$ to the $x$ axis and $\chi(t)$ is the one-axis twisting rate which controls squeezing along the $z$ direction. Squeezing is essential, as it is the only term in the Hamiltonian \eqref{eq:hamiltonian} responsible for the creation of nonclassical states from spin-coherent states~\cite{Kitagawa1993}. A Hamiltonian of this form has been successfully implemented in a variety of experimental settings, including the hyperfine manifolds of cesium atoms~\cite{2007Chaudhury} and dysprosium atoms~\cite{Chalopin2018,2019Nascimbene} as well as in condensates of spin-$1/2$ particles~\cite{2010Leroux,2010Gross,2010Riedel}, and in atomic ensembles in optical cavities where light-mediated interactions induce similar effective spin dynamics~\cite{Li2022}. For strontium atoms, this Hamiltonian has been investigated numerically for the universal generation of quantum states and gates~\cite{2021Omanakuttan}, and has also been realized experimentally~\cite{2024Deutsch}. In addition, the one-axis twisting term $\chi J_z^2$ has already been realized on several experimental platforms~\cite{2002Molmer,2016Kasevich,2016Bohnet}.

We will show that the number of control parameters can be reduced while still allowing the generation of AC states. To do this, we fix $\phi(t) = \pi/2$. The Hamiltonian \eqref{eq:hamiltonian} then reduces to the well-studied one-axis twisting and rotation Hamiltonian $H(t)=\Omega(t)J_{y}+\chi(t) J_{z}^{2}$, which can be used to produce extreme spin squeezed states achieving the Heisenberg limit, see e.g.~\cite{Carrasco2022}. Our protocol exploits this Hamiltonian to maximize the objective function $\mathcal{A}_t(|\psi\rangle\langle\psi|)$ starting from a pure coherent spin state $|\psi_0\rangle$. In QOC, the most efficient algorithms rely on the calculation of the gradient of the objective function with respect to the controls. However, in our case, gradient-based methods, such as LBFGS or gradient descent with the use of automatic differentiation, frequently only found local minima, resulting in a suboptimal set of parameters and anticoherence measure. Initially, to find controls that generate AC states, we used the well-known gradient-free CRAB algorithm~\cite{2022Muller} with the Hamiltonian~\eqref{eq:hamiltonian}, which gave satisfactory results for small orders of anticoherence. For example, we were able to find controls that generate spin states close to anticoherent states to order $5$, with $1-\mathcal{A}_5<10^{-3}$ for $j=9$. However, we developed a pulse-based protocol that uses the same control parameters, specifically designed for the generation of AC states, which greatly outperformed CRAB in terms of speed, convergence and scalability. Despite the limitations of the gradient-free Nelder-Mead method, our approach has enabled us to find controls that can successfully produce higher-order AC states (up to $t=9$ for $j=24$) and that can handle large spins (up to $j=5000$ for $t=2$). Furthermore, because our protocol applies the pulses sequentially, it is not limited by the bandwidth of the control frequency, unlike in QOC.

\begin{figure}
\begin{centering}
\includegraphics[width=0.65\linewidth]{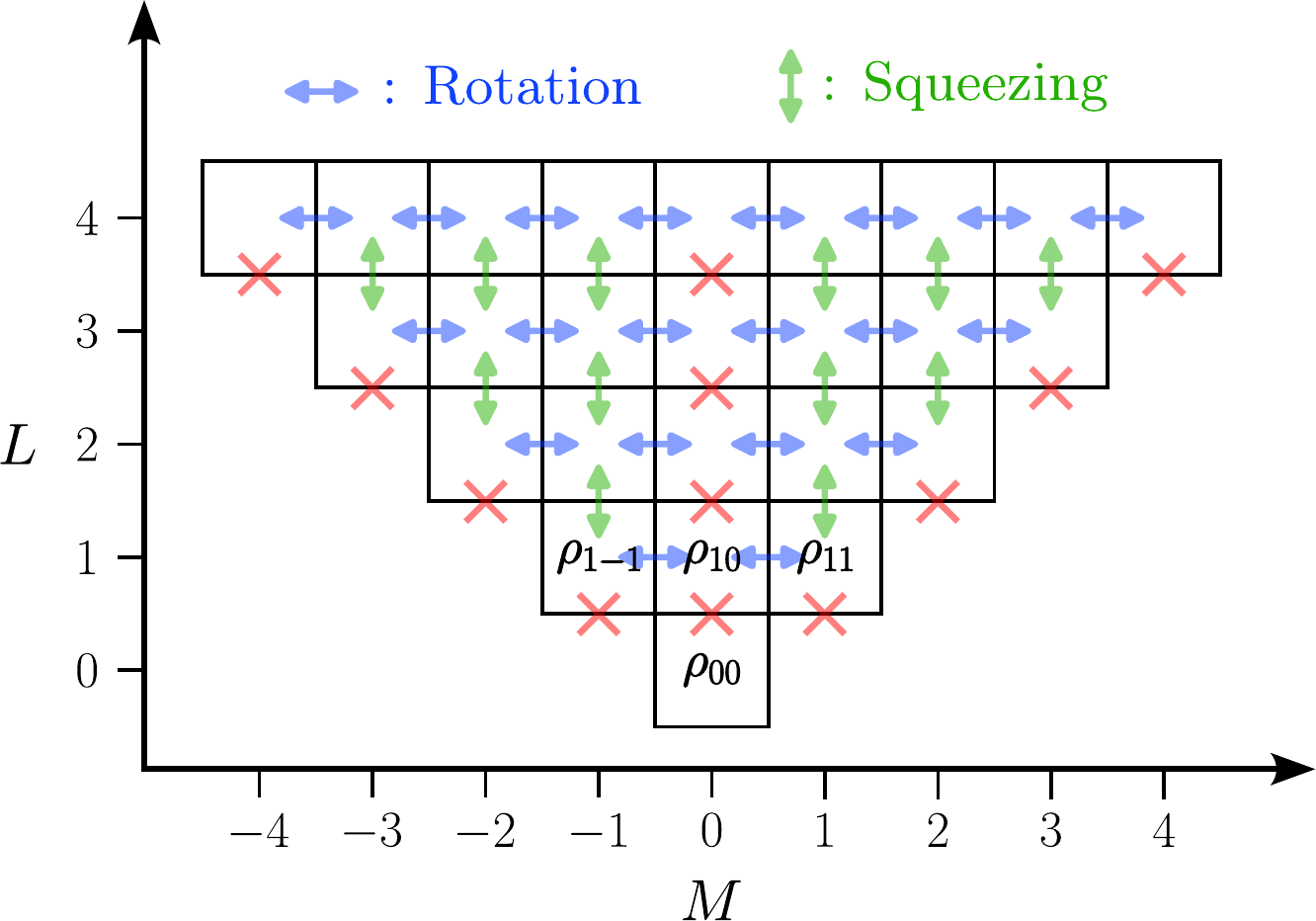}
\par\end{centering}
\caption{Diagram illustrating the coupling between state multipoles (for $j=2$). Each square represents a multipole $\rho_{LM}$ of the spin state from the expansion \eqref{rhoexpansion} (these are shown explicitly for $L=0,1$). Blue and green arrows indicate the effects of rotation ($J_y$ generator) and squeezing ($J_z^2$ generator), respectively. Red crosses denote either the absence of coupling between two multipoles or the absence of an adjacent multipole.\label{fig:Multipoles_coupling}}
\end{figure}

\subsection{Pulse-based protocol}
\label{Sec:pulseprotocol}

The key idea behind our pulse-based protocol lies in the distinct multipole coupling behaviors of the squeezing and rotation operations. Squeezing generated by $J_z^2$ couples a multipole $\rho_{LM}$ only to its neighbors $\rho_{L\pm1,M}$, while the rotation generated by $J_y$ affects only multipoles with the same $L$ (see Fig.~\ref{fig:Multipoles_coupling} and Appendix~\ref{Appendix:MultipolesEvolutionSqueezing} for details). Through squeezing, population can therefore be transferred from a multipole at level $L$ to one at level $L+1$, which is desirable when seeking to generate anticoherent states, in which all multipole moments of order less than or equal to $t$ are suppressed. While squeezing enables this upward transfer, it can also cause reverse coupling from $L+1$ to $L$, which can reintroduce lower-order moments and prevent anticoherence from being achieved. However, this can be avoided by using a rotation that places the $L+1$ multipoles in specific $M$ states, namely $M = 0$ and $M = \pm L$, which are decoupled from the lower levels. By transferring the population to these decoupled states, we can apply a squeezing while keeping those multipoles occupied in the upper level, allowing others in $L$ to move to $L+1$ without unwanted backflow. Conversely, a rotation can also move a lower multipole from a decoupled position ($M=0$ or $M=\pm L$) to a configuration where further squeezing pushes it upward. Figure~\ref{fig:N=6_q=_pulsebased_analytical} clearly demonstrates this behavior; see Section~\ref{subsec:AnalyticalResults}.

Based on this idea, our protocol consists of a sequence of $n_{C}$ cycles, where each cycle (except the first) applies a rotation about the $y$-axis followed by a squeezing operation along $z$. Thus, in the one-axis twisting and rotation Hamiltonian, we alternate between applying squeezing ($\chi(t) \neq 0$, $\Omega(t) = 0$) and rotation ($\Omega(t) \neq 0$, $\chi(t) = 0$) by activating only one term at a time. The corresponding operations are described by the operators
\begin{equation*}
R_{y}(\theta) = e^{-iJ_{y}\theta}, \qquad S_{z}(\eta) = e^{-iJ_{z}^{2}\eta}
\end{equation*}
where $\theta$ and $\eta$ are the amplitudes of the rotation and the squeezing. Note that this sequence of pulses is similar to the protocol presented in~\cite{Carrasco2022}. The final state of the system after $n_C$ cycles is then given by
\begin{equation}
    |\psi_{n_C}\rangle = \left(\prod_{i=1}^{n_C}S_{z}(\eta_i)\,R_{y}(\theta_i)\right)|\psi_0\rangle
\end{equation}
where $\theta_1$ is always taken as zero since it is necessary to first perform a squeezing. The initial state $|\psi_0\rangle$ is the coherent state that points in the direction of the $y$-axis. For each sequence, we optimize the parameters $\{\theta_i,\,i=2,3,\dots,n_C\}$ and $\{\eta_i,\,i=1,2,\dots,n_C\}$ to generate an AC state of a given order, that is, a state that maximizes the AC measure \eqref{eq:bures_acmeasure} for a given $t$. As we shall see, while this approach is fully realizable with the Hamiltonian \eqref{eq:hamiltonian}, it is specifically tailored for AC state generation rather than producing arbitrary spin states. This protocol is experimentally accessible with current technology, both in terms of the necessary gates~\cite{Carrasco2022} and the attainable experimental parameters $\theta$ and $\eta$~\cite{2024Gupta,2020Morello}. For example, the coherence times of the coherent and spin cat states in a Sb donor nucleus ($j=7/2$) implanted in silicon-based chip are respectively of $T\approx100$~ms and $T\approx14$~ms. Based on conservative value of the squeezing strength $\chi$, a typical squeezing parameter $\eta=\chi t=\pi/2$ is achievable in $4.375$~ms~\cite{2024Gupta,2020Morello}, well within the coherence time of the system.

\begin{figure}
\begin{centering}
\includegraphics[width=\linewidth]{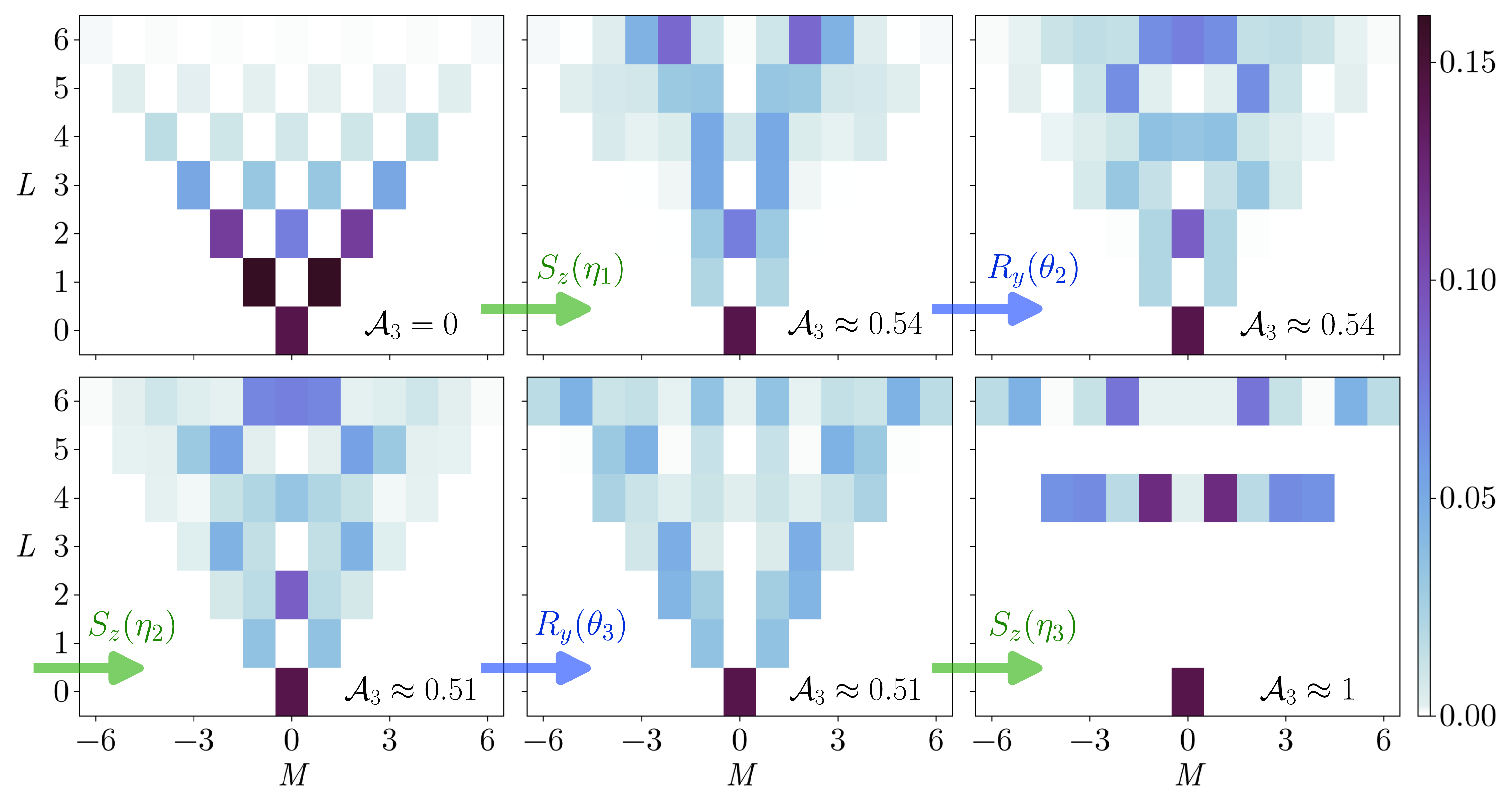}
\par\end{centering}
\caption{Protocol for generating anticoherent states of order $3$ for spin $j = 3$ using $n_C=3$ cycles. The optimization is performed over all parameters ($\eta_1,\eta_2,\eta_3,\theta_2$ and $\theta_3$). Each colored rectangle represents the modulus squared of the corresponding multipole of the expansion \eqref{rhoexpansion}. The final anticoherence measure reaches $1-\mathcal{A}_{3}<10^{-7}$. \label{fig:N=6_q=_pulsebased}}
\end{figure}

\subsection{Results}
\subsubsection{Numerical optimization}
\label{subsec:NumericalOptimization}
We first optimize the parameters $\eta_i$ and $\theta_i$ numerically using the gradient-free Nelder–Mead algorithm. Figure~\ref{fig:N=6_q=_pulsebased} shows the pulse sequence obtained for $j=3$, which prepares an AC state of order $3$ with a deviation $1 - \mathcal{A}_3 < 10^{-7}$ in $n_C = 3$ cycles. The first squeezing operation, $S_z(\eta_1)$, transfers the population from the lower-order multipoles $\rho_{2\pm 2}$ to the higher-order multipoles $\rho_{6\pm 2}$. The subsequent rotation, $R_y(\theta_2)$, shifts the dominant population within $L = 6$ from $M = \pm 2$ to $M = 0$. As discussed previously, the $M = 0$ components are decoupled from lower-order multipoles under squeezing, allowing the next squeezing step to preserve these higher-order contributions without transferring them back. Finally, the rotation $R_y(\theta_3)$ removes any residual population in $\rho_{20}$ by transferring it to $\rho_{2\pm1}$ and $\rho_{2\pm 2}$, which are completely eliminated by the final squeezing operation.

\begin{figure}
\begin{centering}
\includegraphics[width=0.9\linewidth]{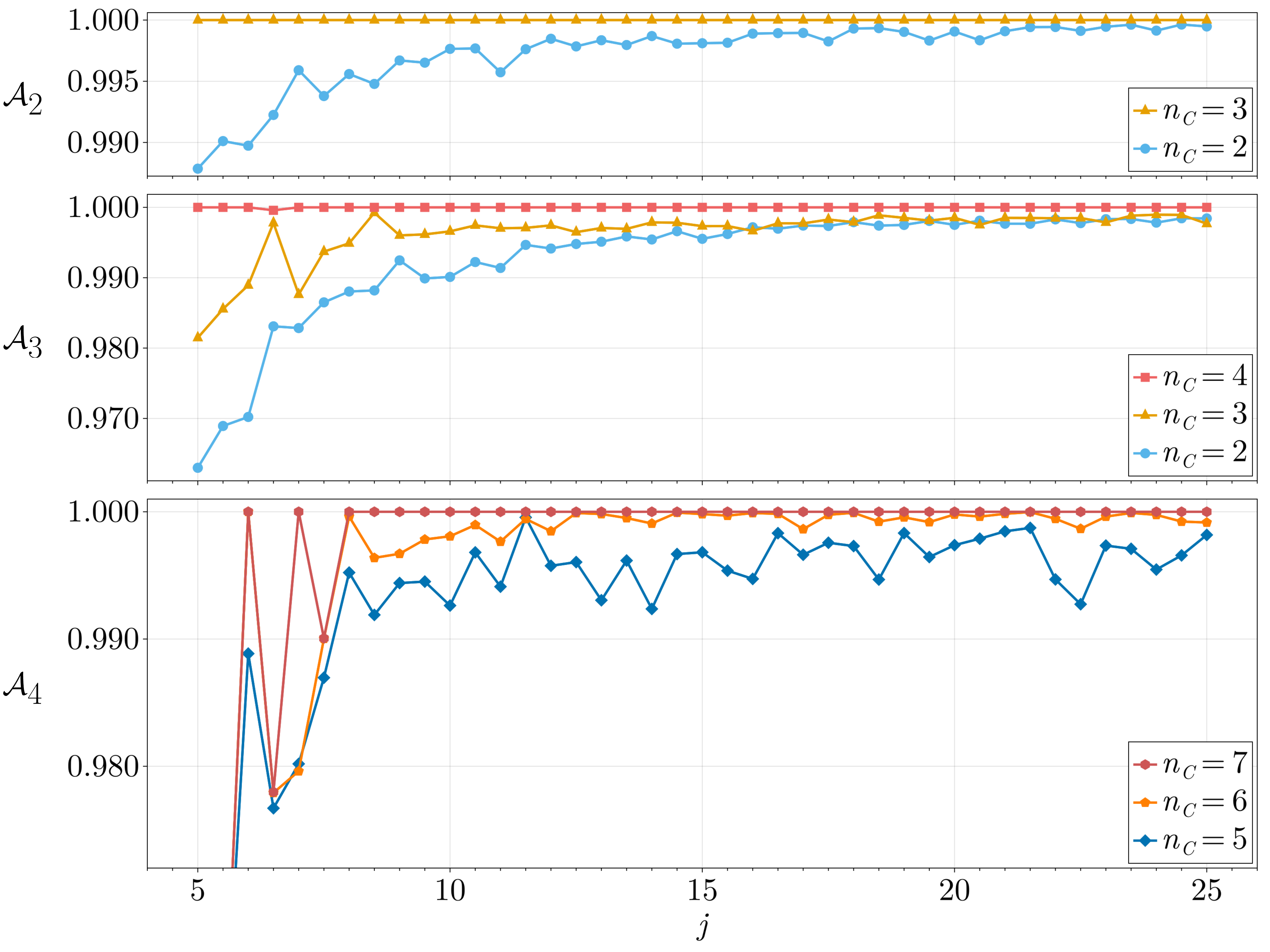}
\par\end{centering}
\caption{Highest anticoherence measure achieved using the pulse-based protocol for orders $t=2,3$ and $4$ (from top to bottom), shown as a function of the spin quantum number $j$ for different numbers of cycles $n_C$. \label{fig:t=4_N_nC}}
\end{figure}

In Figure~\ref{fig:t=4_N_nC}, we present the maximum AC measures of order $t = 2, 3$ and $4$ obtained by optimizing this protocol, as a function of the spin quantum number $j$ for different numbers of rotation-squeezing cycles $n_C$. The top panel reveals a clear qualitative difference between $n_C = 2$ and $n_C = 3$, the latter ensuring that all generated states satisfy $1 - \mathcal{A}_2 < 10^{-7}$. This suggests that $n_C = 3$ acts as a threshold in the pulse-based protocol to achieve AC to order $t = 2$. A similar threshold behavior is observed for $t = 1$ (data not shown), $t = 3$, and $t = 4$, occurring at $n_C = 1$, $n_C = 4$, and $n_C = 7$, respectively, as illustrated in the second and third panels. This seems to indicate that the pulse-based protocol is indeed specifically optimized to generate AC states. Finally, with $n_C=14$, we are able to successfully generate a state with $\mathcal{A}_9 > 0.99$ for $j = 24$, corresponding to the highest AC order achievable with such precision before the number of cycles becomes too large for the Nelder-Mead optimization to remain effective.

In Table~\ref{tab:ACGeneration_OptimisedParameters}, we present the accumulated values of rotation and squeezing obtained for the generation of AC states of order $t=2,3,\dots,7$, up to numerical errors ($1-\mathcal{A}_t<10^{-15}$). These control parameters were obtained to minimize the total squeezing time, which is anticipated to be the limiting factor on the experimental duration of the protocol. The chosen spin number $j$ is systematically the smallest one for which a given order of anticoherence $t$ is theoretically possible. The values found for each parameter (for $j=2,3,6$ and $12$) are provided in a GitHub repository~\cite{ACGithub}, alongside Julia code used to optimise our protocol.

\begin{table}
\begin{centering}
\begin{tabular}{|c|c|@{\hskip 0.25em}|c|c|c|}
\hline 
$j$ & $\phantom{\Big|}t\phantom{\Big|}$ & $n_{C}$ & Total rotation $\sum_{i=1}^{n_C}|\theta_i|$ & Total squeezing $\sum_{i=1}^{n_C}|\eta_i|$\tabularnewline
\hline 
\hline 
2 & 2 & 2 & 0.560 & 1.323\tabularnewline
\hline 
3 & 3 & 3 & 3.824 & 1.325 \tabularnewline
\hline 
6 & 4 & 6 & 9.818 & 0.959 \tabularnewline
\hline 
6 & 5 & 6 & 6.496 & 1.043 \tabularnewline
\hline 
12 & 6, 7 & 12 & 17.812 & 1.953 \tabularnewline
\hline 
\end{tabular}
\par\end{centering}
\caption{Minimum number of cycles $n_C$ required to generate a pure AC state of order $t$ (with $1-\mathcal{A}_t<10^{-15}$) in a spin-$j$ system. The values chosen for $j$ are the smallest that still allow the generation of an AC state to order $t$. For those values, we observe that $n_C$ always coincides with $j$. The last two columns of the table indicate the accumulated values of rotation and squeezing required to generate the state.}
\label{tab:ACGeneration_OptimisedParameters}
\end{table}

\subsubsection{Analytical results for $t=2$}
\label{subsec:AnalyticalResults}

\begin{figure}
\begin{centering}
\includegraphics[width=\linewidth]{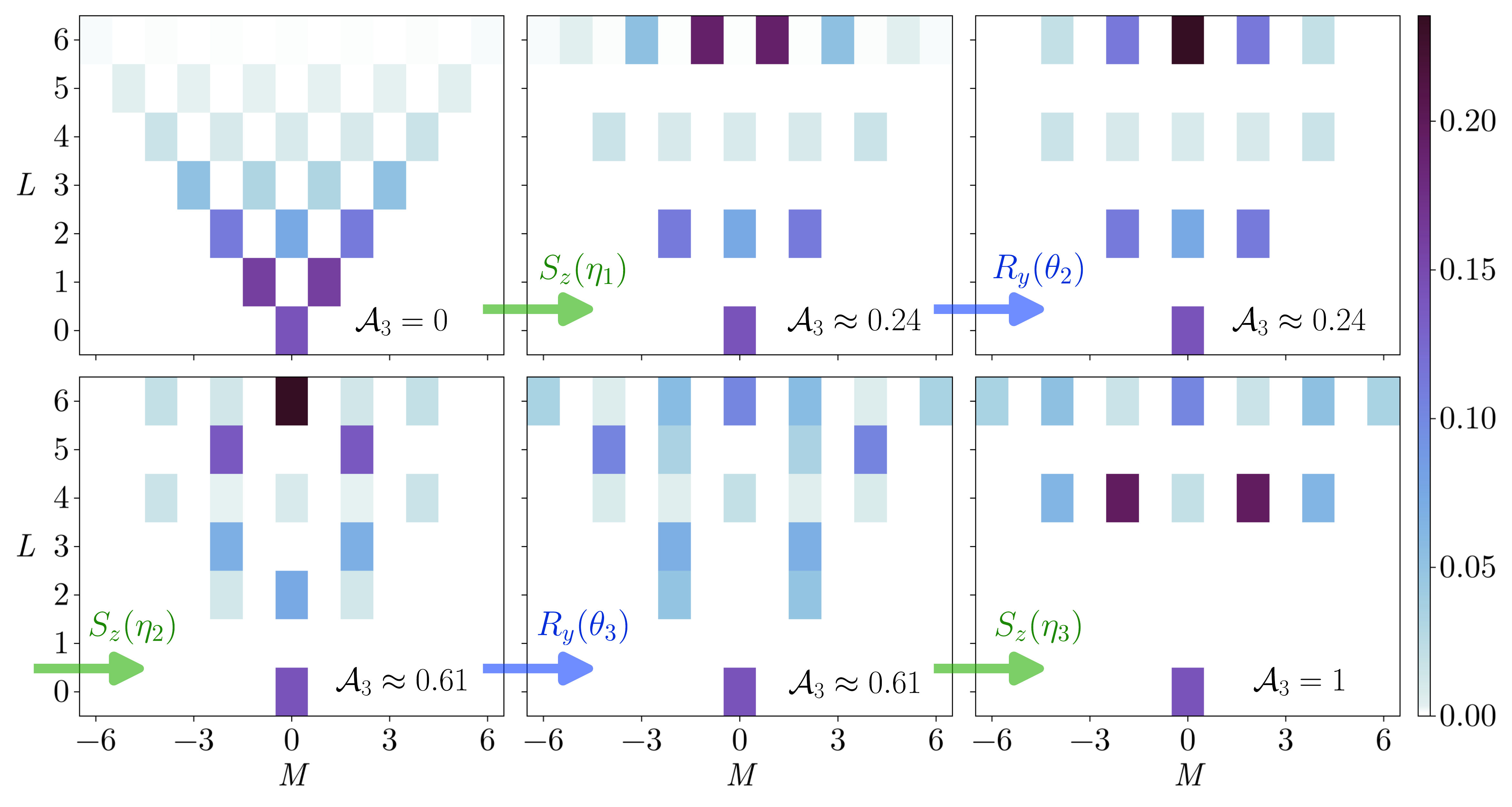}
\par\end{centering}
\caption{Protocol for generating a spin-$3$ AC state of order $3$ based on $n_C=3$ cycles. The control parameters used are those given in Eqs.~\eqref{eq.Rot.AC2} and \eqref{eq.Squ.AC3}.\label{fig:N=6_q=_pulsebased_analytical}}
\end{figure}

Our numerical results show that the cat state, which is AC of order $1$, can always be generated in a single cycle using $\eta_1 = \pi/2$ for any $j$, a finding previously reported and proved in~\cite{1997Agarwal,1999Molmer,2023Cieśliński}.
Similarly, we have just seen in Fig.~\ref{fig:t=4_N_nC} that AC states of order $2$ can be generated from $3$ cycles for all $j$. Based on this observation and on the intuition provided by Fig.~\ref{fig:Multipoles_coupling}, we were able, for integer spin $j$, to derive analytical values for the required control parameters. By examining the effect of the control parameters on the multipoles $T_{LM}$ for $L \leq 2$, we identified that $2$-AC spin-$j$ states could be generated using the following set of squeezing and rotation values
\begin{equation}
    \eta_1=\frac{\pi}{2}, \quad \theta_2=-\frac{\pi}{4j}, \quad \theta_3=\frac{\pi}{2}.\label{eq.Rot.AC2}
\end{equation}
These were subsequently adopted as an ansatz for the next steps of our protocol. Using a symbolic computation program and the results from Appendix \ref{Appendix:MultipolesEvolutionSqueezing}, we further obtained the following squeezing parameters for $j = 2$
\begin{equation}
    \eta_2=-\frac{\mathrm{arccot}\sqrt{2}}{2}, \quad \eta_3=\frac{\mathrm{arccot}\sqrt{2}}{4}.\label{eq.Squ.AC2}
\end{equation}
\begin{figure}[b!]
\centering
\begin{minipage}[c]{0.35\linewidth}
    \centering
    \includegraphics[width=\linewidth]{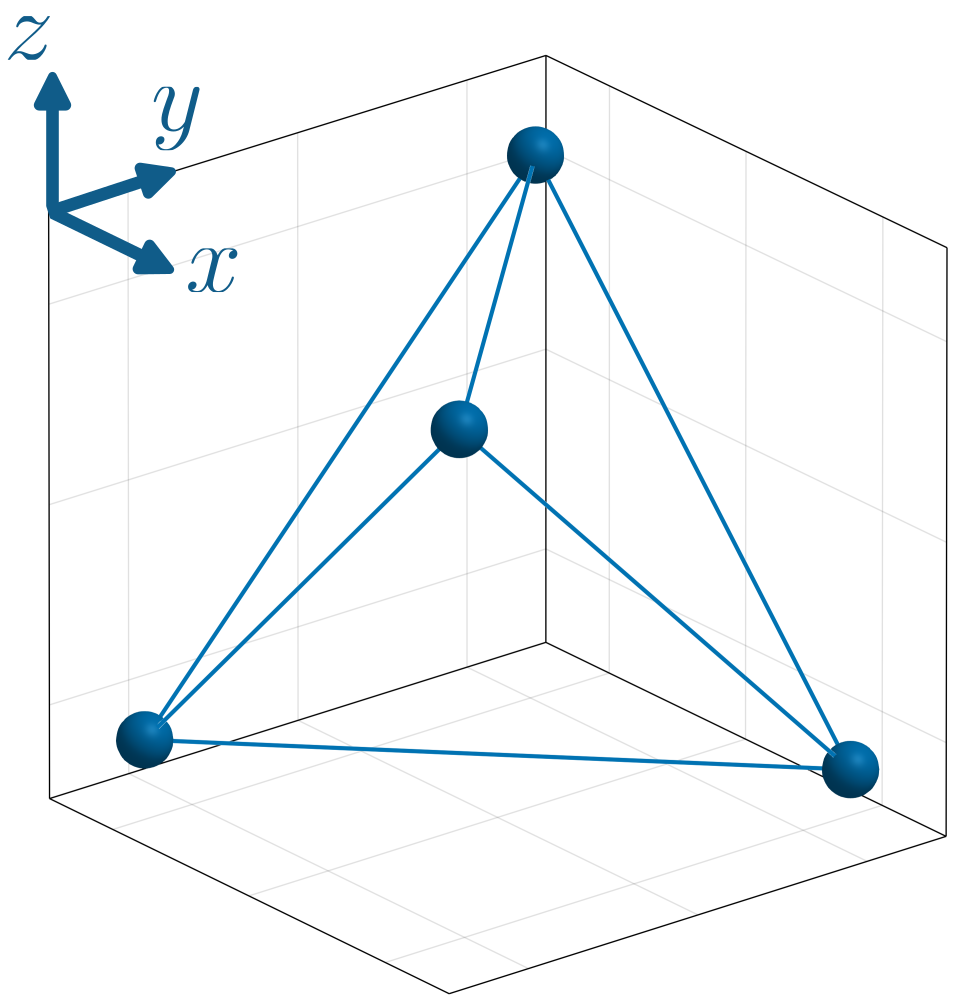}
\end{minipage}
\hspace{0.1\linewidth}
\begin{minipage}[c]{0.35\linewidth}
    \centering
    \includegraphics[width=\linewidth]{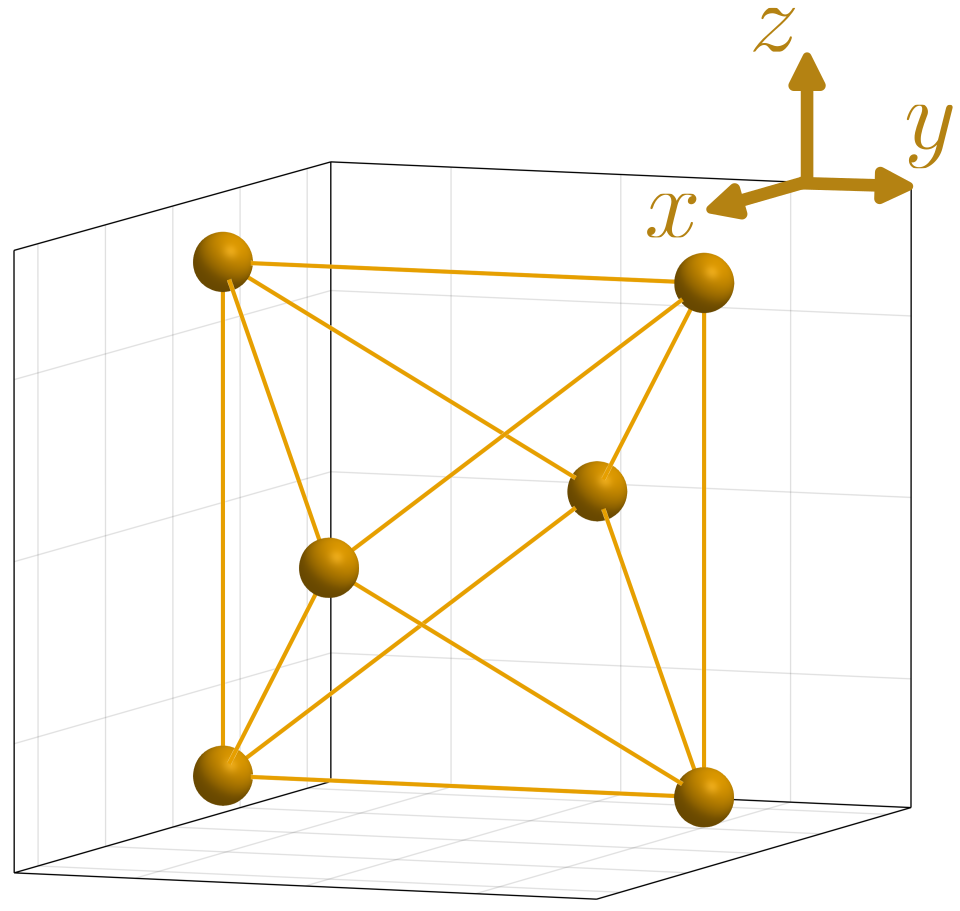}
\end{minipage}
\caption{Majorana representation of the states \eqref{eq:AnalyticalStates_j2} (left) and \eqref{eq:AnalyticalStates_j3} (right) produced by $3$ rotation-squeezing cycles. \label{fig:AnalyticalStates_Majorana}}
\end{figure}
With the control parameters \eqref{eq.Rot.AC2} and \eqref{eq.Squ.AC2}, the generated state is the tetrahedron state (as shown in Fig.~\ref{fig:AnalyticalStates_Majorana})
\begin{equation}
|\psi\rangle=c_1|2,-2\rangle+c_2|2,0\rangle+c_1|2,2\rangle
    \label{eq:AnalyticalStates_j2}
\end{equation}
where 
\begin{equation}
    c_1=\frac{-1/\sqrt{2}+i}{\sqrt{6}},\qquad c_2=\frac{\sqrt{2}+i}{\sqrt{6}}.
\end{equation}
For $j=3$, we found the parameter values
\begin{equation}\label{eq.Squ.AC3}
    \eta_2=-\frac{\mathrm{arccot}\sqrt{2}}{2}, \quad \eta_3=\frac{1}{8}\left[\pi-\arctan\left(2\sqrt{2}\right)\right]
\end{equation}
leading to the octahedron state (also represented in Fig.~\ref{fig:AnalyticalStates_Majorana})
\begin{equation}
    |\psi\rangle=c_1|3,-3\rangle+c_2|3,-1\rangle-c_2|3,1\rangle-c_1|3,3\rangle
        \label{eq:AnalyticalStates_j3}
\end{equation}
where
\begin{equation}
    c_1=-\frac{1}{4} i \left(\frac{1}{3} \left(-241 + 22 \sqrt{2} i \right) \right)^{\frac{1}{8}},
\qquad c_2=-\frac{i \sqrt{5}}{4}\frac{\left(1 + 11  \sqrt{2} i \right)^{1/4}}{3^{5/8}}.
\end{equation}
The latter state is not only AC of order $2$ but also of order $3$. This is a special result, as it is the only AC state of order $3$ that we could obtain with only $n_C=3$ cycles. The other AC states of order $3$ we found needed $n_C=4$ cycles, as can be seen in Fig.~\ref{fig:t=4_N_nC}. We show in Fig.~\ref{fig:N=6_q=_pulsebased_analytical} the evolution of the state multipoles during the generation of the octahedron state \eqref{eq:AnalyticalStates_j3}. It can be compared to the protocol represented in Fig.~\ref{fig:N=6_q=_pulsebased} which also gives an AC state of order $3$ for $j=3$. These analytical results are particularly remarkable because, although the applied controls do not allow the generation of arbitrary spin states, they can still produce exact AC states, as confirmed by our numerical results shown in Fig.~\ref{fig:t=4_N_nC}.

For $j>3$, the parameters $\eta_2$ and $\eta_3$ are determined by numerical optimization to ensure that $1-\mathcal{A}_2<10^{-6}$. For any $j$, the evolution of the multipoles to obtain an AC state of order $2$ from the parameters \eqref{eq.Rot.AC2} is similar to the evolution illustrated for $j=3$ in Fig.~\ref{fig:N=6_q=_pulsebased_analytical}. The process begins by generating the cat state with $\eta_1=\pi/2$. Next, the rotation $\theta_2=-\pi/(4j)$ isolates the highly populated multipoles $\rho_{2j\pm1}$ from the lower levels by transferring them to $\rho_{2j0}$. This is followed by the squeezing $\eta_2$, which shifts the $\rho_{2\pm2}$ multipoles to higher $L$. The subsequent rotation $\theta_3=\frac{\pi}{2}$ fully transfers $\rho_{20}$ to $\rho_{2\pm2}$. Finally, the squeezing $\eta_3$ further moves these multipoles to higher $L$, completing the protocol.

This approach of first generating the $1$-AC state and then the $2$-AC state is not the most time-efficient in terms of squeezing and rotation durations. However, the initial cat state produced by the first squeezing $\eta_1=\pi/2$ could be generated more rapidly using alternative dynamical methods~\cite{2022Huang,2025Dengis,Ansel2022} or based on post-selection~\cite{Alexander2020}, thus reducing the total time required for the spin squeezing. As this first large squeezing $\eta_1=\pi/2$ represents a substantial portion of the total squeezing time, the exploration of alternative methods to generate the cat state could prove to be highly advantageous.

\begin{figure}
\begin{centering}
\includegraphics[width=0.95\linewidth]{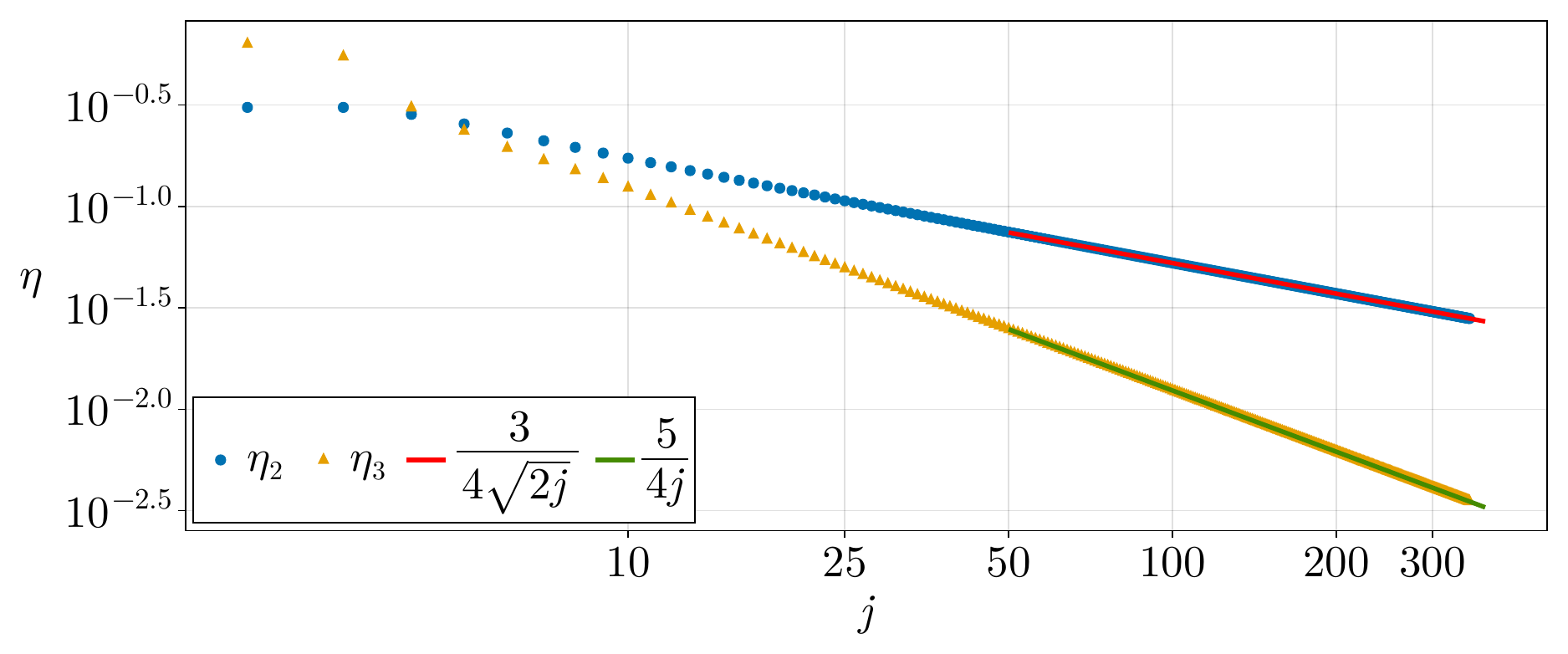}
\par\end{centering}
\caption{Squeezing parameters $\eta_2$ (blue dots) and $\eta_3$ (orange triangles) required for generating a $2$-AC spin-$j$ state with our pulse-based protocol. The other control parameters are set to $\eta_1=\frac{\pi}{2}$, $\theta_1=-\frac{\pi}{4j}$ and $\theta_2=\frac{\pi}{2}$. The red and green lines represent the analytical approximations given in Eq.~\eqref{fiteta}.
\label{fig:squeezingParameters_t=2}}
\end{figure}

Additionally, these analytical values minimize the number of parameters that need optimization, enabling the generation of AC states of order $2$ for larger spin numbers. Figure~\ref{fig:squeezingParameters_t=2} shows the squeezing parameters $\eta_2$ and $\eta_3$, obtained via numerical optimization up to $j = 350$, as functions of $j$ in log-log scale, providing strong evidence that they follow power laws well approximated by
\begin{equation}
    \eta_2(j) = \frac{3}{4\sqrt{2j}}, \qquad \eta_3(j) = \frac{5}{4j}.\label{fiteta}
\end{equation}
The validity of these expressions seems to extend well beyond the fitting region, since by using \eqref{fiteta} for $\eta_2$ and $\eta_3$ for $j=5000$, the generated state has a $2$-AC measure close to $1$ with a deviation $1-\mathcal{A}_2<10^{-3}$.\footnote{For large spin $j$, low-order AC measures of random states sampled according to the Haar measure can be significantly high (see related Ref.~\cite{Goldberg2024}). Therefore, we might suspect that the state generated by our controls for $j = 5000$ is simply a random state with a high $2$-AC measure. However, this is not the case. For a sample of $5000$ random states, we find an average value of $1 - \mathcal{A}_2 > 10^{-2}$ (with a standard deviation $\sigma < 3.3 \cdot 10^{-3}$), which is an order of magnitude larger than that of our generated state.}

\section{Decoherence and mitigation strategies}
\label{Sec.Dec.}

In the previous section, we presented general protocols for generating AC states in quantum systems of total angular momentum $j$, which can be large. These systems may arise from a variety of physical platforms and can represent either individual quantum systems or multipartite systems composed of many particles. Although higher angular momentum enables the generation of higher-order AC states, it also increases their vulnerability to decoherence, as shown in Ref.~\cite{Denis_2022}. Indeed, achieving higher degrees of anticoherence not only accelerates decoherence, but also requires access to larger systems, which are naturally more prone to additional sources of dissipation. Therefore, it is essential to develop strategies aimed at suppressing decoherence and correcting finite-duration pulse errors during the state preparation process in order to preserve, generate, and utilize AC states~\cite{2024Pedram}.

\subsection{Typical sources of dephasing}

\par A practical platform for experimentally realizing the protocols presented in Sec.~\ref{Sec:pulseprotocol} is an ensemble of $N$ spin-$1/2$ particles, where collective spin-$N/2$ states can be generated within the symmetric subspace of the total Hilbert space. In this system, $\mathrm{SU(2)}$ operations correspond to global rotations of the ensemble, while squeezing is achievable, e.g., by coupling the ensemble to a resonant circuit~\cite{boyers_2025}, a mechanical resonator~\cite{Bennett_2013,Xia_2016,Pradana_2021} or a cavity mode~\cite{LewisSwan_2018,Matthew_2018,Pradana_2021}. However, the system is susceptible to coherence loss resulting from undesirable dynamics, such as local disorder and dipolar interactions between neighboring spin-$1/2$. These effects are captured by the error Hamiltonian
\begin{equation}\label{Herr}
    H_{\mathrm{err}} = H_{\mathrm{dis}} + H_{\mathrm{dd}}
\end{equation} with
\begin{equation}
    \begin{aligned}
        H_{\mathrm{dis}} &= \sum_i \delta_i\,\vec{e}_i\bcdot\vec{j}_i,\\
        H_{\mathrm{dd}} &= \sum_{i,j}\Delta_{ij} \qty[3\qty(\vec{e}_{ij}\bcdot\vec{j}_i)\qty(\vec{e}_{ij}\bcdot\vec{j}_j) - \vec{j}_i\bcdot\vec{j}_j]
    \end{aligned}\label{eq:NoRWA}
\end{equation}
where $H_{\mathrm{dis}}$ describes the set of spins $\{\vec{j}_i\}$ rotating around different axes $\qty{\vec{e}_i}$ with different frequencies $\qty{\delta_i}$ and $H_{\mathrm{dd}}$ describes dipole-dipole interactions of frequencies $\qty{\Delta_{ij}}$ between pairs of spins with different orientations. When the Rotating Wave Approximation (RWA) holds, all spins align in the same direction ($z$), leaving only the terms  
\begin{equation}
        H_{\mathrm{dis}} = \sum_i \delta_i j_{i,z},\quad 
        H_{\mathrm{dd}} = \sum_{i,j}\Delta_{ij} [3 j_{i,z}j_{j,z} - \vec{j}_i\bcdot\vec{j}_j].
\label{eq:RWAHamiltonian}
\end{equation}
In the disorder-dominated regime ($\delta_i \gg \Delta_{ij}$), this model captures the main dephasing mechanism to first order in quantum magnetometers based on dense ensembles of NV centers. In this regime, the sensitivity of the sensor is ultimately limited by a combination of factors: strong local disorder resulting from crystal inhomogeneities and interactions with the surrounding spin environment, composed of randomly distributed nuclear spins ($^{13}\mathrm{C}$) and nitrogen defects (P1 centers), as well as dipole interactions between neighboring NV centers~\cite{wolfowicz_2021,zhang_2025,Zhou_2020metrology,Zhou_2023,Taylor_2008}. In such systems, the dipole-dipole Hamiltonian in the rotating frame reads~\cite{Lukin_2017,Zhou_2020metrology} 
\begin{equation}\begin{aligned}
    H_{\mathrm{dd}}^{\mathrm{NV}} &= \sum_{i,j}\Delta_{ij} [2 j_{i,z}j_{j,z} - \vec{j}_i\bcdot\vec{j}_j]\\ 
    & = \sum_{i,j}\Delta_{ij}\qty{\frac{2}{3}[3 j_{i,z}j_{j,z} - \vec{j}_i\bcdot\vec{j}_j] - \frac{1}{3}\vec{j}_i\bcdot\vec{j}_j]}
\end{aligned}\end{equation}
where the Heisenberg Hamiltonian $-\sum_{i,j}\Delta_{ij}\vec{j}_i\bcdot\vec{j}_j$ acts trivially on the collective spin-$N/2$ subspace and hence does not contribute to dephasing to first order of the Magnus expansion. As we will focus solely on first-order decoupling strategies, we may drop the Heisenberg term and use the conventional dipole-dipole Hamiltonian in Eq.~\eqref{eq:RWAHamiltonian}. In the interaction-dominated regime, the model captures the dephasing mechanism in solid-state nuclear spin ensembles, where the dipolar interactions between nuclear spins are typically greater than local disorder due to their low gyromagnetic ratio~\cite{Cappellaro_2022,Lukin_2020,Motte_2016}.

\par Although the dephasing Hamiltonian~\eqref{eq:RWAHamiltonian} generates only unitary dynamics, it does not preserve the polarization of the ensemble and can be described as a form of intrinsic decoherence that induces information leakage out of the collective spin subspace. In this case, the effect of the environment is entirely captured by a static random dephasing term that describes the local magnetic field created by the bath on each spin, introducing disorder into the system. In a more realistic scenario, fluctuations in the local magnetic field due to the dynamics of the bath need to be taken into account, but the static approximation remains valid for time scales much smaller than the memory time of the bath (non zero in the non-Markovian regime). In the case where the bath dynamics is slow compared to the one due to the system-environment coupling, the static approximation is justified to describe the decay of coherence; this is the case, for instance, for an ensemble of NV centers interacting with a spin bath composed of $^{13}\mathrm{C}$ nuclear spins~\cite{Childress_2006} and P1 centers~\cite{DeLange_2010, Hanson_2008,Wang_2013}. It becomes necessary to include the dynamics of the bath in the calculation, for example, when studying the spin-echo decay where low-frequency noise becomes important on the timescale of interest, and this is usually done by replacing the quasi-static dephasing of spin $i$ ($\delta_i$ in Eq.~\eqref{eq:RWAHamiltonian}) with a random, time-dependent dephasing term $\delta_i(t)$ with a correlation function $\expval{\delta_i(t)\delta_i(0)}=\delta_i^2e^{-t/\tau_c}$, associated with a Lorentzian spectral density centered on zero frequency, where $\tau_c$ is the memory time of the bath~\cite{DeLange_2010, Hanson_2008,Wang_2013,Bauch_2020}. In this work, we consider decoupling protocols with total duration $T$ satisfying $T < 1/\sqrt{\langle\delta_i^2\rangle} \ll \tau_c$, justifying the approximation of the system dynamics by a random, static disorder with zero mean and standard deviation $\sqrt{\langle\delta_i^2\rangle}$.

\par Another platform suitable for the generation of AC spin states is based on the spin-$j$ hyperfine manifold of alkali atoms, where rotations can be implemented using a rotating magnetic field, and squeezing is achieved via an off-resonant laser beam~\cite{Deutsch_2010,2021Omanakuttan,Omanakuttan_2024_thesis}. In these systems, decoherence may arise from interactions with a fluctuating magnetic field or from quadrupole interactions with an electric field~\cite{Omanakuttan_2024,onizhuk_2025}. The system dynamics is typically described using the framework of open quantum systems, where decoherence is modeled through a master equation. The dephasing caused by interactions with a magnetic field is represented by a jump operator proportional to the collective spin operator $J_z$, while electric quadrupole interactions are incorporated through the addition of a Hamiltonian term proportional to $J_z^2$, leading to energy level shifts in the spin system. Although the following sections focus only on an interacting spin ensemble, our results apply equally well to the alkali atom platform. This is because the dephasing and quadratic terms $J_z$ and $J_z^2$ transform under the dynamical decoupling sequences considered in this work in the same way as the disorder and dipolar terms in Eq.~\eqref{eq:RWAHamiltonian}.

\par Furthermore, unwanted dynamics that occur during the application of each pulse of the preparation protocol (control pulse), whether rotation or squeezing, can also introduce slight deviations from the intended unitary operations (\ie~finite-duration errors), causing the state preparation protocol to miss the target state. To quantify these finite-duration errors, it is common to move to the so-called \textit{toggling frame}, which is defined as the interaction picture with respect to the control Hamiltonian. For an ideal propagator $U(t)$, which implements a target pulse $U(\tau)=U$ over a time duration $\tau$, the faulty pulse is given by $U_{\mathrm{faulty}} = U e^{-i\tau H_{\mathrm{eff}}}$, where the effective Hamiltonian $H_{\mathrm{eff}}$ generates the finite-duration errors. If decoherence is small enough ($\tau \norm{H_{\mathrm{err}}}\ll 1$ where $\norm{\cdot}$ denotes the operator norm), the effective Hamiltonian can be approximated as
\begin{equation}
    H_{\mathrm{eff}} \approx \frac{1}{\tau}\int_0^{\tau}U^{\dagger}(t)H_{\mathrm{err}}U(t)dt.\label{eq.fin.dur.}
\end{equation}

\subsection{Mitigating dephasing with dynamical decoupling}
In order to mitigate the impact of noise and decoherence in the system, a periodic sequence of pulses, known as a dynamical decoupling (DD) sequence~\cite{Viola_1999, Viola_2013}, can be applied to the system. This DD sequence sequentially and globally rotates the spins of the ensemble in such a way that the system is periodically refocused, preventing errors from accumulating. Many DD sequences have been designed to suppress noise induced by a Hamiltonian of the form \eqref{Herr}, with each sequence showing variable levels of efficiency depending on the specific parameter regime~\cite{Lukin_2020, Cappellaro_2022, Waugh_1968, Cory_1969, Cory_1991}. The choice of the most appropriate sequence depends mainly on the details of the experimental setup, such as the strength of the different noise sources and the minimum pulse duration that can be achieved.

Mitigating finite-duration errors during a control protocol is a more tedious task and requires more advanced techniques, such as the use of \textit{dynamically corrected gates} (DCG)~\cite{DCG_2009_PRA, DCG_2009_PRL, DCG_2010_PRL}. In this scheme, a DD sequence is modified to remove the unwanted Hamiltonian while implementing the intended unitary operation. DCGs were originally constructed to implement simple single- and two-qubit operations in a qubit register, but their design for more complex operations in other quantum systems is not trivial because not all DD sequences are suitable to serve as building blocks for the construction of a DCG. In particular, they must be associated to a \textit{decoupling group} and designed on the \textit{Cayley graph} of that group~\cite{DCG_2009_PRL}, a requirement that is not satisfied by any of the current state-of-the-art sequences for the noise Hamiltonian under consideration~\eqref{eq:RWAHamiltonian}. However, decoupling sequences that satisfy these requirements have recently been introduced and their potential application in the construction of a DCG has been pointed out~\cite{read_2024,Read2025}. In the next section, we use these sequences to construct a DCG that protects the pulse-based protocols described in the previous section from the effect of disorder and dipolar interactions in a spin ensemble.

\section{Robust generation of AC states}
\label{sec:RobustACStates}

In this section, we introduce DCGs designed to perform rotation and squeezing operations protected from finite-duration errors caused by disorder and dipolar interactions in an ensemble of $N$ spin-$1/2$. We then explain how they can be used in the context of protecting the pulse-based protocol described in Sec.~\ref{Sec:pulseprotocol} and demonstrate their effectiveness in the low-decoherence regime of parameters. Finally, we study the effect of control errors on the performance of our DCGs in order to identify the relevant regime of parameters where applying a DCG may improve the fidelity of the control protocol.

\subsection{Dynamically corrected gate design}
\label{sec.DCG.design.}
\begin{figure}[ht]
    \centering
    \includegraphics[width=\linewidth]{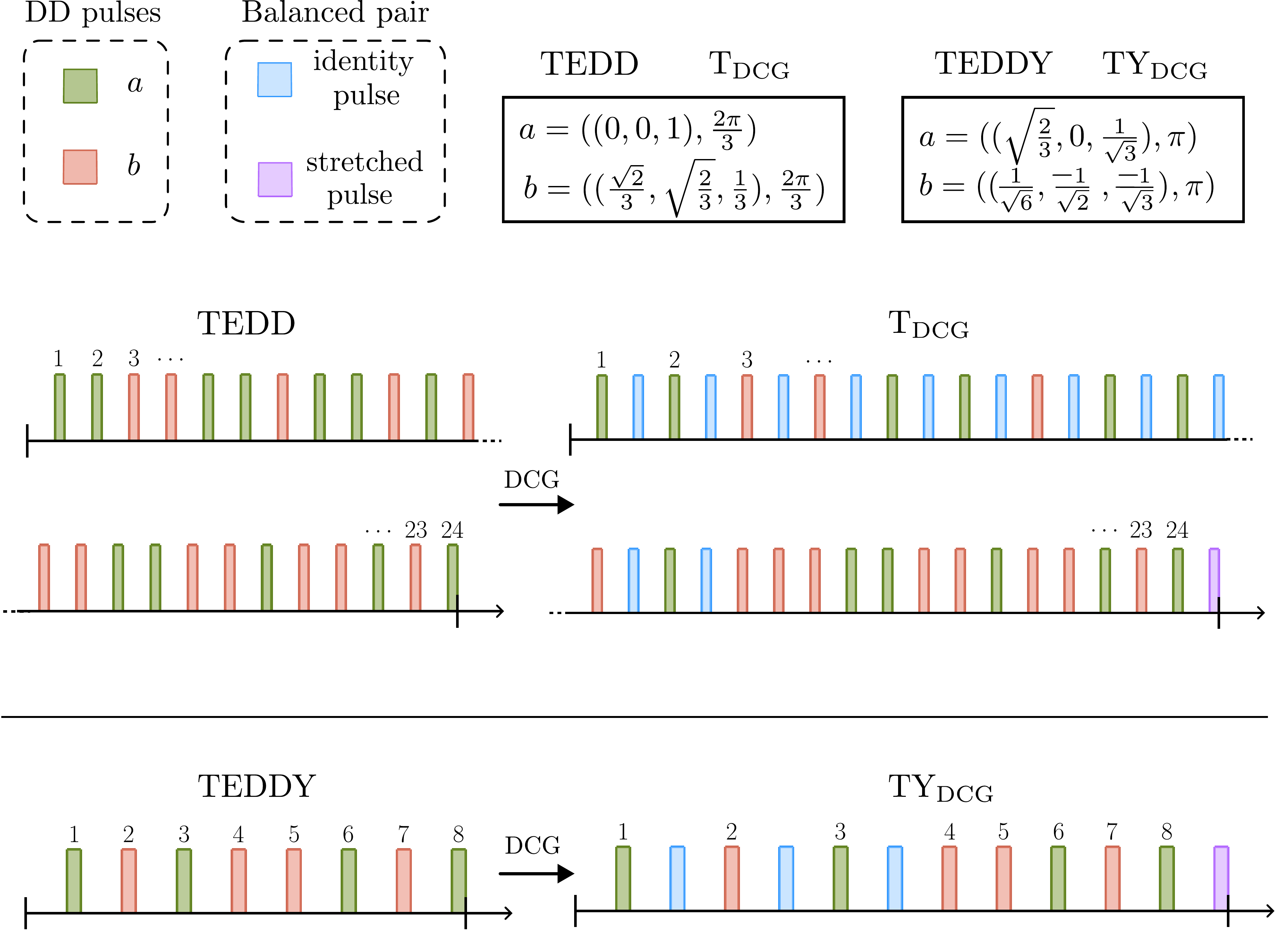}
    \caption{Schematic representation of the DD sequences used in this work (left) and their corresponding dynamically corrected gates (right). In the DCGs, the stretched pulse represents either a rotation unitary, a squeezing unitary or a rotation followed by a squeezing. The DD pulses are labeled using axis-angle notation, as indicated in the boxes.}
    \label{fig_colin_1}
\end{figure}

The building blocks of our two DCGs are the $\mathrm{TEDD}$ and $\mathrm{TEDDY}$ sequences introduced in Refs.~\cite{read_2024} and~\cite{Read2025} and shown in Fig.~\ref{fig_colin_1}. They consist of $24$ and $8$ pulses, each corresponding to one of the two rotations $a$ and $b$ specified in the axis-angle notation for each sequence. $\mathrm{TEDD}$ cancels the general Hamiltonian~\eqref{eq:NoRWA} regardless of whether the RWA holds, while $\mathrm{TEDDY}$ cancels arbitrary disorder but dipolar interaction only under the RWA. We note that the $\mathrm{TEDD}$ sequence considered here slightly differs from the one presented in Ref.~\cite{read_2024}, as we have decided to use an Eulerian path on the Cayley graph that passes by each vertex exactly once during the first half of the sequence. This choice ensures that, in the ideal pulse regime, dephasing is suppressed in the timescale of 12 pulses, instead of 24, which leads to better performances.
To construct the DCGs, we insert $11$ (resp.\ $3$) identity pulses at the appropriate locations in the $\mathrm{TEDD}$ sequence (resp.\ $\mathrm{TEDDY}$), following the procedure described in Refs.~\cite{DCG_2009_PRA, DCG_2009_PRL} and we insert the unitary pulse we want to implement at the end of the sequence (see Fig.~\ref{fig_colin_1}). The resulting sequences, which we call $\mathrm{T_{DCG}}$ and $\mathrm{TY_{DCG}}$, are guaranteed to implement the corresponding unitary operation while refocusing the system, provided that two conditions are met: 
\begin{enumerate}
    \item Firstly, the identity pulse and the target unitary pulse must have the same finite-duration errors. If this is the case, this pair of pulses is called a \textit{balanced pair}. 
    \item Secondly, these finite-duration errors must be suppressed by the DD sequence used in the construction. If the finite-duration error is no longer suppressed by the DD sequence, we say that it has leaked out of the \textit{correctable subspace}, which contains all the Hamiltonians suppressed by the sequence.
\end{enumerate}

We begin by addressing the first condition with a simple construction of balanced pairs. For a simple control Hamiltonian of the form $H(t) = f(t)\, h$, where $h$ is a time-independent operator and $f(t)$ is the control profile that generates a target unitary gate $U(\tau) = U$ over a duration $\tau$, a simple prescription for the balanced pair construction is presented in Ref.~\cite{DCG_2009_PRA}. In this procedure, the unitary pulse at the end of the sequence is obtained by stretching the control profile of the desired gate, such that the new control profile is given by $f_{\mathrm{str}}(t) = \frac{1}{2}f(t/2)$, which implements the same unitary gate $U$ in a duration $2\tau$. The identity pulse is then obtained by applying the intended gate with the original control profile $f(t)$, followed by the reverse time-antisymmetric pulse, which corresponds to a control profile $-f(\tau -t)$. Overall, the balanced pair is defined by the two control profiles
\begin{equation}\label{eq.controlprofile}
\begin{aligned}
    &\text{Stretched pulse profile}:\quad f_{\mathrm{str}}(t) = \frac{1}{2}f(t/2), \quad t \in [0,2\tau] \\[5pt]
     &\text{Identity pulse profile}:\quad f_{\mathrm{id}}(t)  =\begin{cases}
        f(t) \quad &t \in [0,\tau] \\
        -f(2\tau - t) \quad &t \in [\tau, 2\tau]
    \end{cases}
\end{aligned}
\end{equation}
A similar design can be used to construct a balanced pair in a composite pulse sequence. For example, for a composite pulse composed of two successive pulses $U_1(t) = e^{-ih_1\int_0^{t} f_1(t')dt'}$ and $U_2(t) = e^{-ih_2\int_0^{t} f_2(t')dt'}$, the stretched pulse is obtained by individually stretching the profiles $f_1(t)$ and $f_2(t)$. The identity pulse is then obtained by applying the composite pulse, followed by the same pulse sequence executed in reverse, where the reverse time-antisymmetric control profile is used for each pulse. This design therefore requires us to be able to reverse the sign of the control Hamiltonian.

We now turn to the second condition, which requires that the finite-duration error~\eqref{eq.fin.dur.} be effectively suppressed by the DD sequence. Whether this condition is met depends on the specific pulse that is being implemented. For example, when applying a squeezing pulse along the $z$-axis, which commutes with the noise Hamiltonian~\eqref{eq:NoRWA}, the finite-duration errors are indeed suppressed by both DD sequences, and the condition is satisfied. In contrast, for a rotation pulse, the most general form of the finite-duration error takes the form
\begin{equation}
    H_{\mathrm{eff}} = \sum_i\delta_i\, \vec{m}\bcdot \vec{j}_i +\sum_{ij}\Delta_{ij} \qty[3\vec{j}_i\bcdot (M \vec{j}_j) - \vec{j}_i\bcdot\vec{j}_j] 
\end{equation}
where $\vec{m}$ is an unnormalized vector and $M$ a symmetric matrix, both of which depend on the specific rotation being implemented. Their explicit form are given in Appendix~\ref{ap.fin.error.}. In this case, we find that the error remains suppressed by the $\mathrm{TEDD}$ sequence but leaks out of the correctable subspace of $\mathrm{TEDDY}$.

In the case where a composite pulse is to be implemented, finite-duration errors are slightly more complex, and it can be observed that the finite-duration error of a composite pulse consisting of a rotation followed by a squeezing is suppressed, while that of a composite pulse consisting of a squeezing followed by a rotation leaks out of the correctable subspace (see Appendix \ref{ap.fin.error.}).

This leaves us with two options for protecting the pulse-based protocols of Sec~\ref{Sec:pulseprotocol}, which consist of several cycles each consisting of a rotation followed by a squeezing. The first strategy is to protect each cycle as a block using the dynamically corrected gate $\mathrm{T_{DCG}}$, and the second is to protect each rotation individually using $\mathrm{T_{DCG}}$ and each squeezing using $\mathrm{TY_{DCG}}$. 

\subsection{Pulse-based protocols robust to finite-duration errors}
To evaluate the performance of a dynamical decoupling protocol, we use the distance metric~\cite{Lidar_2013,Grace_2010,read_2024}  
\begin{equation}
    D(U,V) = \sqrt{1 - \frac{1}{d_s}\abs{\tr\qty[UV^{\dagger}]}} \label{eq.dist}
\end{equation}  
which quantifies how close a noisy evolution $U$ is to the ideal target unitary $V$ for a system of dimension $d_s$. To identify parameter regimes where incorporating dynamically corrected gates (DCGs) enhances performance, we compute $D(U,V)$ for a pulse-based protocol both with and without DCGs, across a broad range of parameters $(\delta/\chi,\Delta/\chi)$. Here, $\delta=\norm{H_{\mathrm{dis}}}$ and $\Delta = \norm{H_{\mathrm{dd}}}$ denote the strengths of the disorder and dipolar interaction Hamiltonians, respectively, and $\chi$ is the amplitude of the control Hamiltonian. The results are averaged over $20$ randomly generated instances of $H_{\mathrm{dis}}$ and $H_{\mathrm{dd}}$ of the form of Eq.~\eqref{eq:RWAHamiltonian}, using rectangular-shaped control pulses.

\begin{figure}
    \centering
    \includegraphics[width=0.95\linewidth]{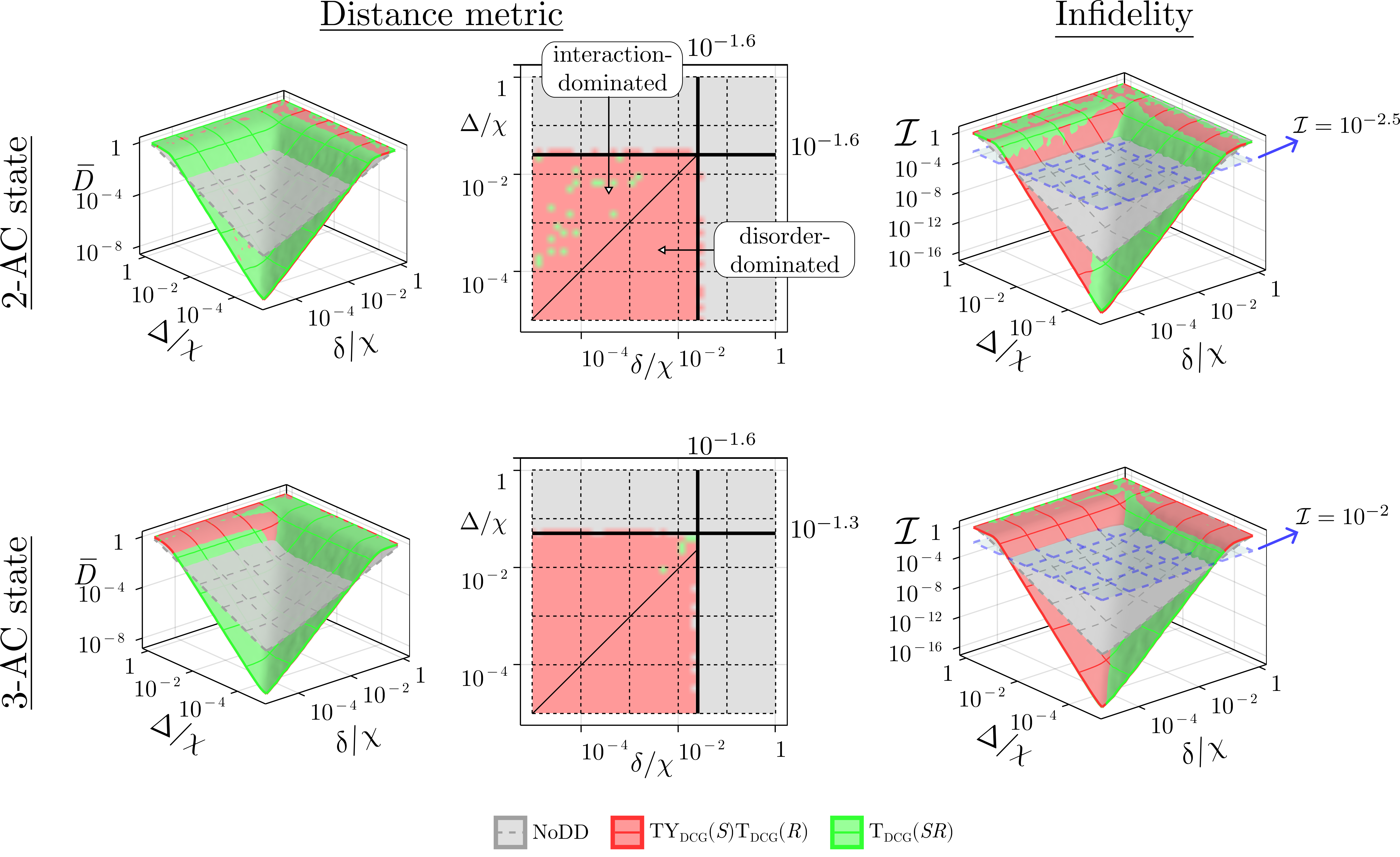}
    \caption{Left: average distance~\eqref{eq.dist} in the $(\delta/\chi,\Delta/\chi)$ parameter space between the ideal and noisy control protocols for the generation of an AC state of order $2$ (top) and $3$ (bottom). Right: infidelity of the state preparation protocol for an AC state of order $2$ (top) and $3$ (bottom). Results are shown for protocols without dynamical decoupling (gray), with DCGs applied to each pulse (red), and with DCGs applied to each cycle (green). The blue line in the center plots, corresponding to $\delta = \Delta$, separates the interaction dominated regime ($\Delta > \delta$) from the disorder-dominated ($\delta > \Delta$) regime.}
    \label{fig_colin2}
\end{figure}

We calculate the distance~\eqref{eq.dist} between the noisy and ideal state preparation protocol propagators of an AC state to order $2$ in a system of $N = 4$ spin-$1/2$ particles, and of an AC state to order $3$ in a system of $N = 6$ spin-$1/2$ particles, using the optimized rotation and squeezing parameters presented in the GitHub repository~\cite{ACGithub} for $t=2$ and $t=3$ respectively. The results are shown in Fig.~\ref{fig_colin2}. We consider two strategies: one where each pulse is individually protected by a DCG (red surface), and another where each cycle is protected as a block (green surface). Both protocols yield smaller distances compared to the unprotected noisy protocol (the 'NoDD' scenario), provided the ratio between the pulse amplitude and the noise amplitude is sufficiently small, as expected~\cite{DCG_2009_PRL}. When this ratio exceeds a critical threshold, the decoupling protocol becomes less effective and introduces more errors than it corrects, due to the significantly longer implementation time of the DCG (see Appendix~\ref{Ap.DCg.An} for more details).

In the absence of pulse imperfections, both DCG strategies perform similarly, as indicated by the near-complete overlap of the red and green surfaces in Fig.~\ref{fig_colin2}. However, protecting each pulse individually results in a slightly smaller distance in all parameter regimes, which can be explained by the superior performance of $\mathrm{TEDDY}$, which cancels out certain higher-order terms, compared to $\mathrm{TEDD}$. In the disorder-dominated regime ($\delta > \Delta$), we find that this strategy outperforms the unprotected protocol (NoDD) when $\delta/\chi\lesssim 10^{-1.6}$ for both the $2$-AC and $3$-AC state preparation protocols. Conversely, in the interaction-dominated regime ($\delta < \Delta$), the performance improvement occurs when $\Delta/\chi \lesssim 10^{-1.6}$ for the $2$-AC protocol and $\Delta/\chi \lesssim 10^{-1.3}$ for the $3$-AC protocol.

\par We may also calculate the infidelity of the prepared quantum state, 
\begin{equation}
    \mathcal{I} = 1-\abs{\bra{\psi_0}V^{\dagger}U\ket{\psi_0}}^2,
\end{equation}
where $\ket{\psi_0}$ is the initial state. We see that the regime where the DCG strategies perform better than NoDD occurs when the infidelity is already quite low, with $\mathcal{I}\lesssim 10^{-2.5}$ (resp. $\mathcal{I}\lesssim 10^{-2}$) for the $2$-AC (resp.\ $3$-AC) state preparation protocol. The DCG protocols should therefore be used in systems where the state-preparation protocol achieves a fidelity already quite high ($\mathcal{F}\gtrsim 99-99.9\%$) and where the leading source of error is decoherence and is not induced by the control pulses. In such systems, dynamically corrected gates can help decrease infidelity by further orders of magnitude~; for instance, for the $2$-AC state preparation protocol, if the initial infidelity is as low as $\mathcal{I}\sim 10^{-3.5}$, the protection offered by the DCG further reduces the infidelity down to $\mathcal{I}\sim 10^{-5}$. We also note that the smallest infidelity is always achieved by the individual pulse protection (resp.\ cycle protection) in the disorder-dominated (resp.\ interaction-dominated) regime, when no control error is taken into account.

\par Such low infidelities are essential, for example, for realizing fault-tolerant quantum computation with logical qubits encoded in a realistic number of physical qubits. In particular, recent works have proposed encoding a logical qubit using spin-$j$ AC states~\cite{Gross_2021,Omanakuttan_2023,JDThesis}. Our protocols could therefore find applications in quantum computing platforms based on atomic or molecular spins, where highly non-classical spin-$j$ states serve as logical qubit encodings~\cite{Asaad_2020, Chizzini_2022, Omanakuttan_2024, Omanakuttan_2024_thesis,Gross_2024,lim_2025}.

\par In order to further increase the efficiency of the protection protocol and its usefulness in experimental setups, it might be possible to optimize the DCG scheme, using the many degrees of freedom in the choice of the pulse sequence, so that the crossover between the NoDD and DCG surfaces in Fig.~\ref{fig_colin2} occurs at greater infidelities and for the DCG to offer better protection over a wider range of parameters. Increasing the amplitude of the DD pulses (red and green pulses in Fig.~\ref{fig_colin_1}) can also enhance performance by reducing the total duration of the DCGs. Although stronger pulses generally introduce larger control errors, we show in the next section that systematic errors in the DD pulses are corrected to first order and are therefore less detrimental. Consequently, using stronger but potentially more error-prone DD pulses can actually be beneficial for maximizing the overall fidelity of the protocol.

\subsection{Effect of control errors}

Dynamically corrected gates are constructed to provide some protection against finite-duration errors, at the cost of a significant control overhead. In particular, the DCGs presented in this work replace a single pulse with a sequence of $36$ pulses for $\mathrm{T_{DCG}}$ and $12$ pulses for $\mathrm{TY_{DCG}}$. This overhead can be detrimental to the system if the errors associated with the pulses, which are not all corrected by the DCG, are non-negligible, in which case the pulse sequence can induce more errors than it corrects. In Appendix~\ref{Ap.C.Error}, we study the effect of control errors on the performance of the DCG strategies presented above, using simple error models consisting of flip-angle errors~\cite{Lidar_2023_survey}, where each rotation or squeezing is subject to a slight over-rotation or over-squeezing. These deviations arise from imperfections in the pulse amplitude, captured by an error parameter $\epsilon$, leading to an effective pulse amplitude of $\chi_{\mathrm{faulty}} = \chi(1+\epsilon)$ rather than the intended $\chi$.

Two distinct types of pulse error must be considered in a DCG: those affecting the DD pulses and those affecting the balanced pair. Errors of the first type (those associated with DD pulses) are automatically corrected to first order by the DCG, provided they are systematic and lie within the correctable subspace of the sequence. This robustness arises from the Eulerian design of the DCG~\cite{Viola_2003, Viola_2013}. In the sequences investigated in this work, correctable errors include systematic over- or under-rotations and axis misspecification, where a rotation is performed around a slightly incorrect axis~\cite{read_2024}. As these errors are self-corrected at first order by the DD sequence, they are less detrimental to the overall performance of the DCG, and we find that a DCG outperforms the unprotected protocol when the control error magnitude satisfies $\epsilon \lesssim \lambda \sqrt{\|H_{\mathrm{err}}\|/\chi}$, where $\|H_{\mathrm{err}}\|$ denotes the supremum operator norm of the noise Hamiltonian~\eqref{eq:RWAHamiltonian}, and $ \lambda$ is a constant that depends on the ratio between the duration of the DCG and of the unprotected protocol, as well as the norm of the second-order term of the Magnus expansion (see Appendix~\ref{Ap.C.Error}). The parameter $\lambda$ can be estimated using analytical upper bounds of the DCG's distance, and we find that $\lambda \sim \sqrt{2/\gamma\alpha^2}$ where $\gamma = \chi \tau$ and $\alpha = \tau_{\mathrm{DCG}}/\tau$ and where $\tau$ (resp.\ $\tau_{\mathrm{DCG}}$) is the duration of the NoDD (resp.\ DCG) protocol. For the GHZ state preparation protocol considered in Appendix~\ref{Ap.C.Error}, we find $\lambda \sim 10^{-1}$. We also note that the performance of the DCG is limited by the control errors in the regime $\epsilon \gtrsim \norm{H_{\mathrm{err}}}/\chi$, in that increasing the amplitude $\chi$ to shorten the DCG duration no longer decreases the distance between the noisy gate and the desired gate, since $\epsilon$ is the leading error parameter.

\par
\begin{figure}[h]
    \centering
    \includegraphics[width=\linewidth]{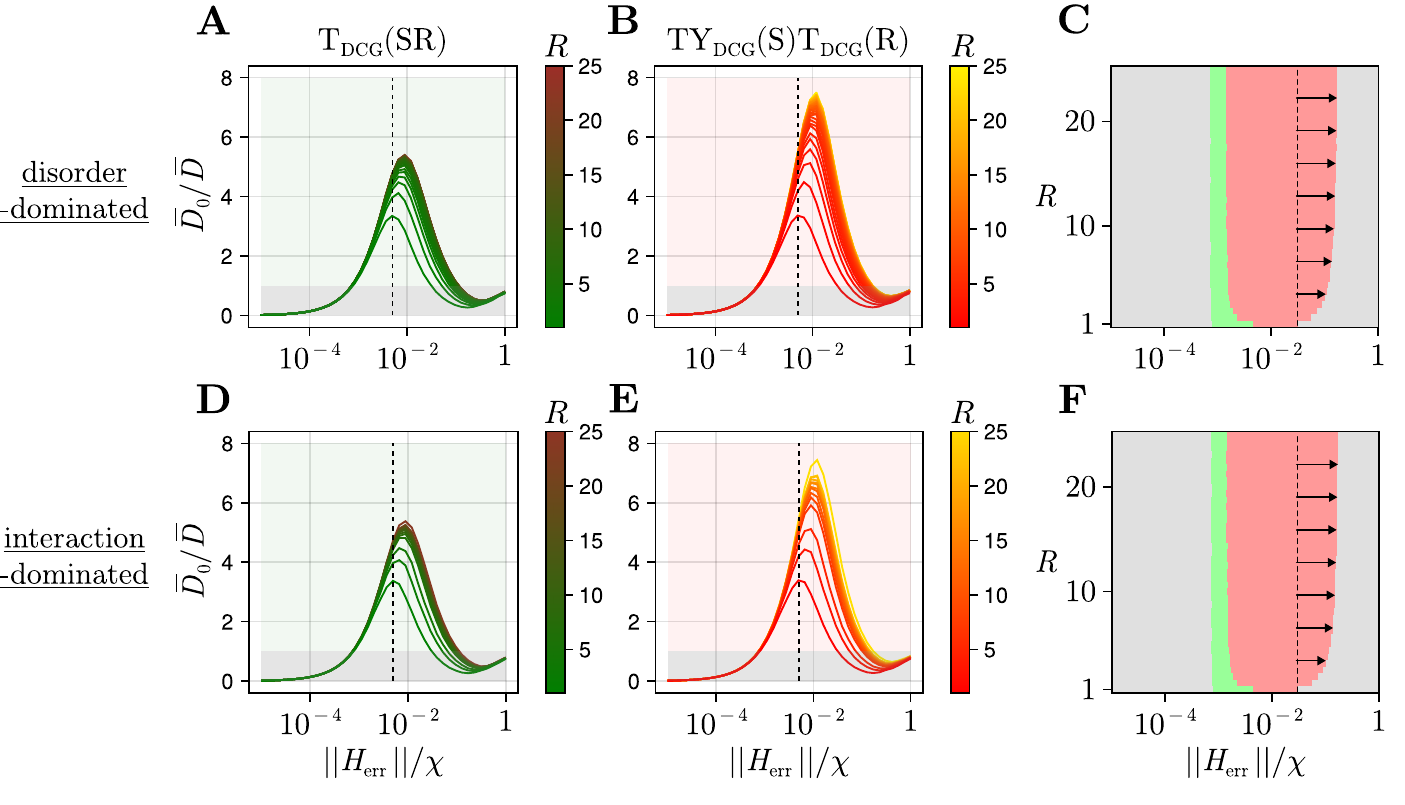}
    \caption{Ratio between the average distances of the unprotected and protected $2$-AC state preparation protocols, shown for the disorder-dominated (A,B) and interaction-dominated (D,E) regimes, and for the cycle (A,D) and individual-pulse (B,E) protection schemes.
    The ratio is plotted as a function of the noise strength $\norm{H_{\mathrm{err}}}/\chi$ for various values of $R$. The dashed line in panels (A,B,D,E) indicates the condition $\norm{H_{\mathrm{err}}}/\chi = \epsilon$. Panels (C,F) display heatmaps identifying the best-performing protocol in the ($\norm{H_{\mathrm{err}}}/\chi$,$R$) space, using the same color code as in Fig.~\ref{fig_colin2}. The black arrows highlight the extension of the DCG performance regime with increasing $R$.}
    \label{fig:NewFig}
\end{figure}

 As discussed in the previous section, the self-correction inherent to the Eulerian design allows one to enhance the performance of the DCG by increasing the amplitude of the DD pulses (denoted $\chi_{\mathrm{DD}}$). This reduces the total duration of the protocol, albeit at the cost of additional errors arising from the stronger pulses. However, the resulting fidelity improvement becomes negligible once the finite-duration errors of the DD pulses are small compared to the finite-duration errors of the balanced pair or to the control errors of the DD pulses themselves. When $\epsilon$ is independent of $\chi_{\mathrm{DD}}$, a simple error analysis (see Appendix~\ref{Ap.DCg.An}) reveals that increasing $R \equiv \chi_{\mathrm{DD}}/\chi$ further does not significantly improve the fidelity once 
\begin{equation}
    1 \ll R\qty(\frac{\theta_{\mathrm{BP}}}{\theta_{\mathrm{DD}}} + \frac{\epsilon \chi}{\norm{H_{\mathrm{err}}}}),
\end{equation}
where $\theta_{\mathrm{DD}}$ (resp.\ $\theta_{\mathrm{BP}}$) denotes the sum of the absolute values of all rotation angles implemented by the DD pulses (resp.\ the rotation angles and squeezing parameters implemented by the balanced pairs). For the $2$-AC state preparation protocol, where $\theta_{\mathrm{BP}}/\theta_{\mathrm{DD}} \approx 1.7$ (resp.\ $0.4$) when the squeezing and rotation pulses are protected as a cycle (resp.\ individually), we expect that increasing $R$ will favor the $\mathrm{TY_{DCG}(S)T_{DCG}(R)}$ protocol. This protocol reduces the parameter $\theta_{\mathrm{BP}}$, which determines the total duration required to implement all balanced pairs. The ratio $\overline{D}_0/\overline{D}$ between the average distances of the unprotected and protected protocols is plotted in Fig.~\ref{fig:NewFig} for various values of $R$ and $\norm{H_{\mathrm{err}}}/\chi$. Results are shown for the disorder-dominated regime ($\delta/\Delta = 10$, panels A,B) and the interaction-dominated regime ($\delta/\Delta = 0.1$, panels D,E), considering both cycle protection  (panels A,D) and individual pulse protection (panels B,E), with a fixed flip-angle error parameter $\epsilon = 0.5\%$. A ratio $\overline{D}_0/\overline{D}>1$ (resp.\ $<1$) indicates an improvement (resp.\ no improvement) over the unprotected NoDD protocol corresponding to a reduction of the infidelity by a factor proportional to $(\overline{D}_0/\overline{D})^2$. In the regime $\epsilon \gg \norm{H_{\mathrm{err}}}/\chi$, the performance of both protocols is largely independent of $R$, since $\epsilon \chi/\norm{H_{\mathrm{err}}}$ is already much greater than one. Conversely, when $\epsilon < \norm{H_{\mathrm{err}}}/\chi$, performance improves significantly with increasing $R$, as shown by the extension of the parameter region where the DCG protocol outperforms NoDD, and by a several-fold increase in the ratio $\overline{D}_0/\overline{D}$. Finally, the heatmaps in panels C,F identify which protocol performs best as a function of $R$ and $\norm{H_{\mathrm{err}}}/\chi$, demonstrating the widening of the high-performance regime of the DCG with increasing $R$.

\begin{table}[t]
    \centering
    \renewcommand{\arraystretch}{1.6} 
    \setcellgapes{5pt} 
    \makegapedcells 
    \begin{tabular}{|c|c|c|c|}
        \cline{2-4}
        \multicolumn{1}{c|}{} & \multirow{2}{*}{Errors in DD pulses} & \multicolumn{2}{c|}{Errors in Balanced Pair} \\\cline{3-4}
        \multicolumn{1}{c|}{} &  & Type I & Type II \\\cline{1-4}
        DCG better than NoDD & $\epsilon\lesssim \lambda \sqrt{\frac{\norm{H_{\mathrm{err}}}}{\chi}}$&$\epsilon\lesssim \lambda \sqrt{\frac{\norm{H_{\mathrm{err}}}}{\chi}}$  & $\epsilon\lesssim \frac{\norm{H_{\mathrm{err}}}}{\chi}$ \\ \cline{1-4}
        DCG performance bound & \makecell[cc]{$\frac{\norm{H_{\mathrm{err}}}}{\chi} \lesssim \epsilon$} & \makecell[cc]{$\frac{\norm{H_{\mathrm{err}}}}{\chi} \lesssim \frac{1}{\lambda}\sqrt{\epsilon}$} & \makecell[cc]{$\frac{\norm{H_{\mathrm{err}}}}{\chi} \lesssim \frac{1}{\beta \lambda}\sqrt{\epsilon}$} \\
        \hline
    \end{tabular}
    \caption{Performance regimes of a DCG when control errors are introduced in the pulse sequence. We consider errors in the DD pulses and in the balanced pair, distinguishing between type I errors (where identity pulses introduce no additional errors but fail to correct over-rotation) and type II errors (where identity pulses imperfectly implement the identity, introducing new errors). We find the parameter regimes where DCG outperforms the NoDD protocol and where its performance is limited by control errors.}
    \label{tab:my_label}
\end{table}

\par Unfortunately, the errors associated with the balanced pair are typically not corrected and may more easily accumulate to significantly reduce the performance of the DCG, or even prevent any benefit. Flip-angle errors in the balanced pair are modeled by their effect on the control profile of the stretched and identity pulses, and we replace the profiles given in~\eqref{eq.controlprofile} by 
\begin{equation}\begin{aligned}
    & \text{NoDD profile}:\quad  f_{\mathrm{NoDD}}(t) = (1+\epsilon)f(t), \quad t \in [0,\tau] \\[5pt]
    & \text{Stretched pulse profile}:\quad f_{\mathrm{str}}(t) = \frac{1}{2}(1+\epsilon^{\mathrm{str}})f(t/2), \quad t \in [0,2\tau]\\[5pt]
     &\text{Identity pulse profile}: \quad f_{\mathrm{id}}(t) =\begin{cases}
        (1+\epsilon)f(t) \quad &t \in [0,\tau] \\
        -(1+\epsilon^{\mathrm{id}})f(2\tau - t) \quad &t \in [\tau, 2\tau]
    \end{cases}
\end{aligned}\end{equation}
The error parameters $\epsilon$, $\epsilon^{\mathrm{str}}$ and $\epsilon^{\mathrm{id}}$ can in general differ, as the control error can vary if the control profile is stretched or the sign of the Hamiltonian is reversed. We find that two different types of balanced pair errors have very different impacts on the DCG performance, which can be modeled by choosing different sets of parameters $(\epsilon, \epsilon^{\mathrm{str}},\epsilon^{\mathrm{id}})$. For $\epsilon = \epsilon^{\mathrm{str}}=\epsilon^{\mathrm{id}}$, each identity pulse implements the identity operation perfectly and introduces no additional errors into the system, so that the DCG implements the error-prone pulse without introducing any additional errors, but without correcting the over-rotation. We refer to these errors as type I errors. In this case, we find that the DCG still outperforms the NoDD protocol in the regime $\epsilon \lesssim \lambda \sqrt{\norm{H_{\mathrm{err}}}/\chi}$, but that DCG's performance is now limited by the control errors in the regime where $\norm{H_{\mathrm{err}}}/\chi \lesssim \lambda \sqrt{\epsilon}$.

\par Finally, it should be noted that other types of control error can result in additional errors introduced by each identity pulse, which is the case when $\epsilon = \epsilon^{\mathrm{str}}=-\epsilon^{\mathrm{id}}$. In this particular case, each identity pulse in the DCG implements an operation that differs slightly from the identity due to $\epsilon$, and the DCG not only fails to correct the control error but also introduces new ones~; we refer to these errors as type II errors. This extends the regime in which DCG provides no improvement over NoDD to $\epsilon \gtrsim \norm{H_{\mathrm{err}}}/\chi$, and the performance of the DCG remains limited by control errors in the regime where $\norm{H_{\mathrm{err}}}/\chi \lesssim \gamma \sqrt{\epsilon}$, with $\gamma \sim \sqrt{\beta}\lambda$ and $\beta = \tau_{\mathrm{BP}}/\tau$ being the ratio between the time spent on the balanced pair and the duration of the NoDD protocol. Moreover, in the presence of such errors and in the regime $\epsilon \lesssim \norm{H_{\mathrm{err}}}/\chi \lesssim \gamma \sqrt{\epsilon}$, the DCG strategy where each pulse is protected individually performs much better than the strategy where each cycle is protected as a block, because the $\mathrm{TY_{DCG}}$ sequence requires fewer identity pulses and thus introduces fewer errors into the system. This strategy should therefore be considered when errors during the squeezing pulses are non-negligible and reducing the total squeezing time is crucial to the outcome of the control protocol.

\section{Conclusion}
\label{sec:Conclusion}

In this work, we have presented a simple and efficient protocol for the deterministic generation and protection of anticoherent (AC) spin states using a combination of spin rotation and squeezing operations. Our pulse-based protocol, which involves cycles of rotations followed by squeezing, has demonstrated remarkable efficiency in producing AC states of different orders, achieving a high degree of anticoherence even for large spin quantum numbers. Through numerical optimization and analytical derivations, we have identified the optimum parameters for the rotation and squeezing operations, enabling the generation of AC states up to order 9 for spin-24 systems and order 2 for collective spin ensembles.

To address the inherent fragility of AC states to decoherence, we developed dynamically corrected gates (DCGs) capable of implementing the pulse-based state preparation protocol while suppressing the relevant noise mechanisms. We have shown that our methods effectively mitigate dephasing arising from dipole-dipole interactions and on-site disorder in interacting spin ensembles, preserving coherence during state preparation by our protocol. Our analysis demonstrates that single-pulse protection using DCGs outperforms full-cycle protection in disorder-dominated regimes, while the latter performs very similarly in the interaction-dominated scenarios. Furthermore, we have shown that DCGs remain advantageous when control errors are sufficiently small and that the self-correcting properties of the sequences with respect to certain pulse errors can be leveraged to improve the sequence performance, supporting the feasibility of experimentally producing  $t$-AC states for various orders $t$. These states hold great promise for applications in quantum sensing, metrology, and fundamental quantum studies, where their rotational invariance and sensitivity to perturbations can be exploited.

Our protocols can be applied to any physical platform where rotation and squeezing operations are possible, such as magnetic atoms, spin ensembles, or even Bose-Einstein condensates. Future research could then explore the use of more sophisticated Hamiltonian dynamics, such as two-axis anisotropic countertwisting~\cite{2023Witkowska} or effective three-body collective-spin interactions~\cite{2024Zhang}, to further speed up the generation of AC states. Beyond deterministic coherent control schemes, AC states could also be generated probabilistically via post-selection or through dissipative state preparation methods. An interesting direction to explore could be the use of quantum non-demolition measurement schemes based on multicolor probing, as demonstrated in~\cite{Saffman2009}, which may offer another approach to generate AC states in atomic ensembles with increased robustness against technical noise and inhomogeneous broadening. Investigating these alternatives may offer new avenues for producing practically useful AC states under less stringent coherence requirements, expanding their applicability to realistic experimental platforms.

\section*{Acknowledgements}
We thank Eduardo Serrano-Ens\'astiga for fruitful discussions.

\paragraph{Author contributions}
JM initiated and supervised the project throughout. JD established the AC states production protocols, while CR developed the dynamical decoupling techniques. All authors actively participated in discussions throughout the project and contributed to the writing of the manuscript.

\paragraph{Funding information}
JM acknowledges the FWO and the F.R.S.-FNRS for their funding as part of the Excellence of Science program (EOS project 40007526). CR is a Research Fellow of the F.R.S.-FNRS. Computational resources were provided by the Consortium des Equipements de Calcul Intensif (CECI), funded by the Fonds de la Recherche Scientifique de Belgique (F.R.S.-FNRS) under Grant No.~2.5020.11.

\begin{appendix}
\numberwithin{equation}{section}

\section{Evolution of multipoles under rotation and squeezing}
\label{Appendix:MultipolesEvolutionSqueezing}
In this Appendix, we show that the rotation generator $J_y$ couples a multipole of order $M$ only to those of order $M \pm 1$, without changing the value of $L$, while the squeezing generator $J_z^2$ couples a multipole of order $L$ only to those of order $L \pm 1$, without altering the value of $M$. Throughout this section, we set $\hbar=1$.

\subsection{Rotation}
Under the unitary evolution generated by the Hamiltonian $\Omega J_y$, the density matrix in the multipolar basis evolves according to 
\begin{equation}
i\sum_{LM}\dot{\rho}_{LM}T_{LM}=\Omega\sum_{LM}\rho_{LM}\left[J_y,T_{LM}\right].  \label{eq:multipolarrotationevolution}
\end{equation}
The operator $J_y$ can be expressed in terms of the ladder operators as
\begin{equation}
    J_y = \frac{J_+-J_-}{2i}
\end{equation}
and its commutator with any multipole operator is given by
\begin{equation}
[J_\pm,T_{LM}]=\sqrt{L(L+1)}C_{LM,1\pm1}^{LM\pm1}T_{LM\pm1}.
\end{equation}
Therefore, the commutator in Eq.~\eqref{eq:multipolarrotationevolution} can be rewritten in the form
\begin{equation}
    \left[J_y,T_{LM}\right] = \frac{\sqrt{L(L+1)}}{2i}\left(C_{LM,11}^{LM+1}T_{LM+1}-C_{LM,1-1}^{LM-1}T_{LM-1}\right).
\end{equation}
Using the relations
\begin{equation}
    C_{LM,1-1}^{LM-1}=\frac{\sqrt{(L-M+1)(L+M)}}{\sqrt{2L(L+1)}},\qquad C_{LM,11}^{LM+1}=-\frac{\sqrt{(L+M+1)(L-M)}}{\sqrt{2L(L+1)}}
\end{equation}
and the orthogonality relation
\begin{equation}
    \mathrm{Tr}\left(T_{LM}T_{L'M'}^{\dagger}\right) = \delta_{LL'}\delta_{MM'}, \label{eq:TLM_orthogonality}
\end{equation}
we find that the evolution of any multipole component $\rho_{LM}$ is governed by
\begin{equation}
    \sum_{LM}\dot{\rho}_{LM}=\frac{\Omega}{2\sqrt{2}}\big[(L-M+1)(L+M)\rho_{LM-1}+(L+M+1)(L-M)\rho_{LM+1}\big],
\end{equation}
which depends only on the neighboring multipoles $\rho_{LM\pm1}$.

\subsection{Squeezing}
Under the unitary evolution generated by the Hamiltonian $\chi J_{z}^{2}$, the density matrix in the multipolar basis evolves according to 
\begin{equation}
i\sum_{LM}\dot{\rho}_{LM}T_{LM}=\chi\sum_{LM}\rho_{LM}\left[J_{z}^{2},T_{LM}\right].  \label{eq:multipolarsqueezingevolution}
\end{equation}
In the multipolar basis, the squeezing operator is given by 
\begin{equation}
J_{z}^{2} = \frac{j(j+1)\sqrt{2j+1}}{3}T_{00}+\frac{1}{6\sqrt{5}}\sqrt{\frac{(2j+3)!}{(2j-2)!}}T_{20}.
\end{equation}
Since $ T_{00} $ is proportional to the identity matrix, it does not contribute to the evolution of the density matrix, meaning only the term involving $ T_{20} $ need to be considered. We can now use the general commutator between two multipolar operators, given by~\cite{Var.Mos.Khe:88}
\begin{multline}
    \left[T_{L_{1}M_{1}},T_{L_{2}M_{2}}\right] = \sqrt{(2L_{1}+1)(2L_{2}+1)}\;\sum_{L}(-1)^{2j+L} \left(1 - (-1)^{L_{1}+L_{2}+L} \right)\\
    \times \begin{Bmatrix} L_{1} & L_{2} & L \\ j & j & j \end{Bmatrix} C_{L_{1}M_{1},L_{2}M_{2}}^{LM_{1}+M_{2}} T_{LM_{1}+M_{2}},
\end{multline}
where we used the $ 6j $-symbol and Clebsch-Gordan coefficients. These coefficients are non-zero only when $ |L_{1} - L_{2}| \leq L \leq L_{1} + L_{2} $. In our case, since $ L_{1} = 2 $, the maximum multipolar order reachable from $ L_{2} $ is $ L = L_{2} \pm 2 $. However, for $ L = L_{2} \pm 2 $ or $ L = L_{2} $, the factor  
\begin{equation}
1 - (-1)^{L_{1} + L_{2} + L}
\end{equation}
vanishes, leading to the final expression
\begin{multline}
    \left[J_{z}^{2},T_{LM}\right] = \frac{M}{\sqrt{2L+1}} \left( \sqrt{\frac{(L-M+1)(L+M+1)(2j-L)(2j+L+2)}{2L+3}} T_{L+1M} \right.\\
    + \left. \sqrt{\frac{(L-M)(L+M)(2j-L+1)(2j+L+1)}{2L-1}} T_{L-1M} \right).
\end{multline}
Finally, using the orthogonality property \eqref{eq:TLM_orthogonality}, Eq.~\eqref{eq:multipolarsqueezingevolution} simplifies to
\begin{multline}
    \dot{\rho}_{LM} = \frac{\chi}{i} \frac{M}{\sqrt{2L+1}} \left( \sqrt{\frac{(L-M)(L+M)(2j-L+1)(2j+L+1)}{2L-1}} \rho_{L-1M} \right.\\
    + \left. \sqrt{\frac{(L-M+1)(L+M+1)(2j-L)(2j+L+2)}{2L+3}} \rho_{L+1M} \right),
\end{multline}
which clearly shows that a multipole of order $L$ is coupled only to its adjacent multipoles of order $ L \pm 1 $.

\section{Finite-duration errors and leakage out of the correctable subspace}
\label{ap.fin.error.}
Consider a unitary evolution operator $U(t)$, implementing a target unitary $U(\tau)=U$ in a time $\tau$, and the noise Hamiltonian 
\begin{equation}
    H_{\mathrm{err}} = \sum_i\delta_i j_{i,z} + \sum_{ij}\Delta_{ij}\qty[3 j_{i,z}j_{j,z} - \vec{j}_i\bcdot \vec{j}_j]
\end{equation}
which induces finite-duration errors. The error can be quantified by moving to the toggling frame with respect to $U(t)$, where the noise Hamiltonian reads, in the case where $\tau \norm{H_{\mathrm{err}}} \ll 1$,
\begin{equation}
    H_{\mathrm{eff}} \approx \frac{1}{\tau}\int_0^{\tau} dt\,U^{\dagger}(t)H_{\mathrm{err}}U(t).
\end{equation}
We aim to determine whether the effective Hamiltonian $H_{\mathrm{eff}}$ leaks out of the correctable subspace of the $\mathrm{TEDD}$ and $\mathrm{TEDDY}$ sequences for several relevant propagators, including elementary squeezing and rotation pulses, as well as composite pulses consisting of squeezing and rotation.

\subsection{Elementary pulses}

\paragraph{Squeezing} In the case of a squeezing pulse, $U(t) = e^{-i\chi(t)J_z^2}$ (with $J_z=\sum_i j_{i,z}$ the collective spin) and we have $[U(t),H_{\mathrm{err}}]=0\,\forall t$, so that $H_{\mathrm{eff}} = H_{\mathrm{err}}$ and no leakage occurs.

\paragraph{Rotation} In the case of a rotation $R[\vec{n}(t),\theta(t)] \equiv R(t)\in\mathrm{SO(3)}$, $ U(t) = e^{-i\theta(t) \hat{n}(t)\bcdot \vec{J}}$ and we have
\begin{equation}
    H_{\mathrm{eff}} = \sum_i\delta_i \vec{m}\bcdot \vec{j}_i +\sum_{ij}\Delta_{ij} \qty[3\vec{j}_i\bcdot (M\vec{j}_j)  - \vec{j}_i\bcdot\vec{j}_j]\label{eq.FD.form} 
\end{equation}
where $\vec{m} = \frac{1}{\tau}\int_0^{\tau}dt R(t)\vec{z}$ is an unnormalized vector and $M$ is a $3\times 3$ symmetric matrix whose entries are given by $M_{ij} = \frac{1}{\tau}\int_0^{\tau} dt\, R_{zi}(t)R_{zj}(t)$. While the disorder term is still in the correctable subspace as for squeezing, the dipolar term leaks out of the correctable subspace of $\mathrm{TEDDY}$ which corrects only two-body interactions proportional to $3j_{i,z}j_{j,z} - \vec{j}_i\bcdot\vec{j}_j$. On the other hand, $\mathrm{TEDD}$ corrects all two-body interactions written as $H_{ij} = \sum_{\alpha,\beta}h_{\alpha\beta}^{ij}j_{\alpha,i}j_{\beta,i}$ that satisfy $\tr\qty[h^{ij}]=0$. Indeed, in this case, it is easy to verify that 
\begin{equation}\begin{aligned}
    \tr \qty[3M - \mathds{1}_{3\times 3}] = 3\frac{1}{\tau}\int_0^{\tau} dt\, \qty[R_{zx}^2(t) + R_{zy}^2(t) + R_{zz}^2(t)] -3 = 3\frac{1}{\tau}\int_0^{\tau} dt -3= 0
\end{aligned}\end{equation}
using the property that $R(t)$ is an $\mathrm{SO}(3)$ rotation matrix with unit-norm rows and columns at all times. Consequently, $\mathrm{TEDD}$ suppresses the Hamiltonian~\eqref{eq.FD.form}.

\subsection{Composite pulses}

Consider a composite pulse composed of two pulses $U_1(t)$ and $U_2(t)$, implementing the unitaries $U_1(\tau_1)=U_1$ and $U_2(\tau_2)=U_2$ in a time duration $\tau_1$ and $\tau_2$ respectively. The finite-duration error of the composite pulse can be written as $H_{\mathrm{eff}} = H_{\mathrm{eff}}^{(1)} + H_{\mathrm{eff}}^{(2)}$ with 
\begin{equation}\begin{aligned}
    H_{\mathrm{eff}}^{(1)} &= \frac{1}{\tau_1 + \tau_2} \int_0^{\tau_1 }dt U_1^{\dagger}(t)H_{\mathrm{err}} U_1(t), \\ 
     H_{\mathrm{eff}}^{(2)} &= \frac{1}{\tau_1 + \tau_2} U_1^{\dagger}\qty[\int_0^{\tau_2 }dt U_2^{\dagger}(t)H_{\mathrm{err}} U_2(t)]U_1,
\end{aligned}\end{equation}
where $H_{\mathrm{eff}}^{(1)}$ and $H_{\mathrm{eff}}^{(2)}$ represent the finite-duration errors of the first and second pulses, respectively, with $H_{\mathrm{eff}}^{(2)}$ evaluated assuming that $U_1$ has already been applied.

\par In the case where $U_1(t)$ is a rotation and $U_2(t)$ is a squeezing along the $z$ axis, the total finite-duration error can be suppressed by $\mathrm{TEDD}$, as both $H_{\mathrm{eff}}^{(1)}$ and $H_{\mathrm{eff}}^{(2)}$ can be written in the form~\eqref{eq.FD.form}. However, when $U_1(t)$ is a squeezing and $U_2(t)$ is a rotation, squeezing $U_1$ does not commute with $U_2^{\dagger}(t) H_{\mathrm{err}} U_2(t)$ at all times. As a result, the finite-duration error $H_{\mathrm{eff}}^{(2)}$ cannot be decomposed into simple disorder and dipolar terms. Instead, it includes more complex $K$-body interactions that are not corrected by our sequence.

\section{Effect of control errors on the DCG}\label{Ap.C.Error}

\begin{figure}[ht]
    \centering
    \includegraphics[width=0.7\linewidth]{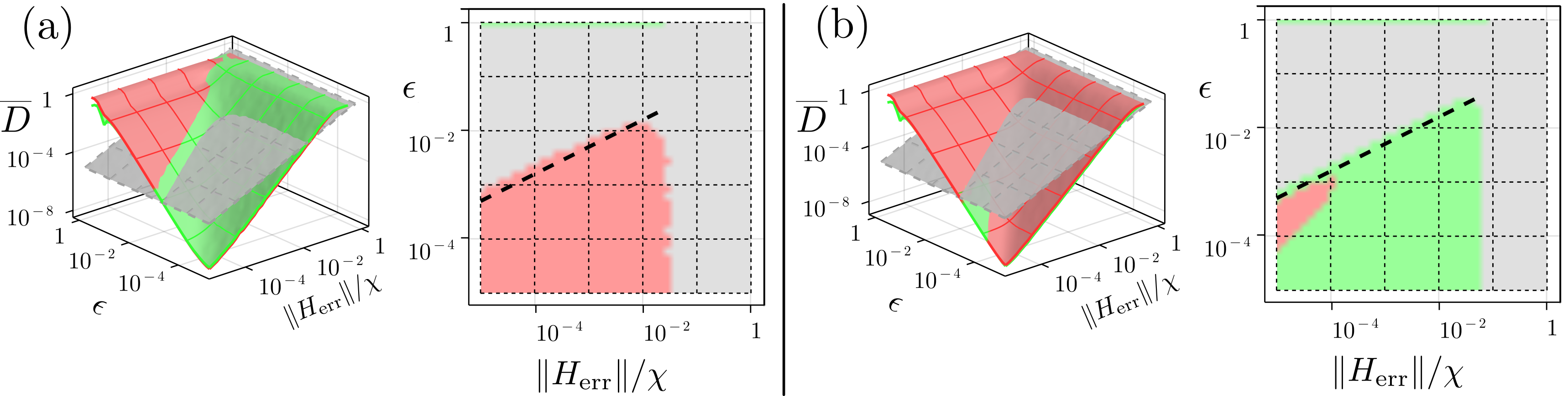}
    \caption{Average distance between the ideal and noisy control protocol propagators for the preparation of a GHZ state in an ensemble of $4$ spin-$1/2$ under the different DCG strategies.}
    \label{fig.GHZ}
\end{figure}

We study the effect of control errors on the performance of our pulse-based protocol protected from finite-duration errors by DCGs. For that purpose, we consider the simplest protocol presented in this work, which prepares a GHZ state starting from a coherent state pointing in the direction of the $z$-axis by applying a $\pi/2$ rotation about the $x$-axis followed by a squeezing of strength $\pi/2$ along the $z$-axis. For an ensemble of $N=4$ spin-$1/2$, we calculate the distance~\eqref{eq.dist} between the noisy and ideal state-preparation propagators for a wide range of parameters $(\delta/\chi,\Delta/\chi)$, where $\delta=\norm{H_{\mathrm{dis}}}$ and $\Delta = \norm{H_{\mathrm{dd}}}$, in the NoDD case, when the rotation and squeezing pulses are individually protected by a DCG and when the whole cycle is protected by a single DCG (results shown in Fig.~\ref{fig.GHZ}). In the absence of control errors, we find that protecting each pulse individually performs better than protecting a cycle in the disorder-dominated regime, while the cycle protection scheme works best in the interaction-dominated regime. In general, the best DCG strategy only provides an improvement over NoDD in the parameter regimes $\delta/\chi\lesssim 10^{-1.6}$ and $\Delta/\chi\lesssim 10^{-1.3}$.

\par We then introduce control errors in the protocols and study how they impact the DCG's performance by calculating the distance measure in the $(\epsilon,\norm{H_{\mathrm{err}}}/\chi)$ parameter space, where $\norm{H_{\mathrm{err}}} = \norm{H_{\mathrm{dis}} + H_{\mathrm{dd}}}$, considering both the interaction- and disorder-dominated regime by fixing $\norm{H_{\mathrm{dis}}}/\norm{H_{\mathrm{dd}}} = 10^{-1}$ and $\norm{H_{\mathrm{dd}}}/\norm{H_{\mathrm{dis}}} = 10^{-1}$ respectively. We consider three types of error, namely (i) errors in the DD pulses, (ii) errors in the balanced pair for which the identity pulses do not introduce additional errors and (iii) errors in the balanced pair where each identity pulse introduces errors to the system. As a simple model for control errors, we consider flip-angle errors~\cite{Lidar_2023_survey} (over- or under- rotations) where the amplitude of a faulty pulse is given by $\chi_{\mathrm{faulty}} = \chi(1+\epsilon)$ where $\abs{\epsilon}\ll 1$. For convenience, we concentrate on the case where $\epsilon>0$, which generates over-rotations.

\subsection{Errors in the DD pulses}

Let us consider that the amplitude of each DD pulse slightly deviates from the intended amplitude by a small error parameter $\epsilon\ll 1$, which systematically over-rotate each spin, and that the balanced pairs and the NoDD protocol are error-free. To study the effect of these errors on our protocols, we construct an AC state of order $1$ in an ensemble of $4$ spin-$1/2$, using the NoDD and the DCGs strategies and calculate the distance~\eqref{eq.dist} in the $(\norm{H_{\mathrm{err}}}/\chi, \epsilon)$ parameter space, where $\norm{H_{\mathrm{err}}} = \norm{H_{\mathrm{dis}} + H_{\mathrm{dd}}}$ is the norm of the Hamiltonian~\eqref{eq:RWAHamiltonian}. We consider the disorder-dominated and interaction-dominated regimes by fixing $\norm{H_{\mathrm{dd}}}/\norm{H_{\mathrm{dis}}} = 10^{-1}$ and $\norm{H_{\mathrm{dis}}}/\norm{H_{\mathrm{dd}}} = 10^{-1}$ respectively. 

\begin{figure}[ht]
    \centering
    \includegraphics[width=\linewidth]{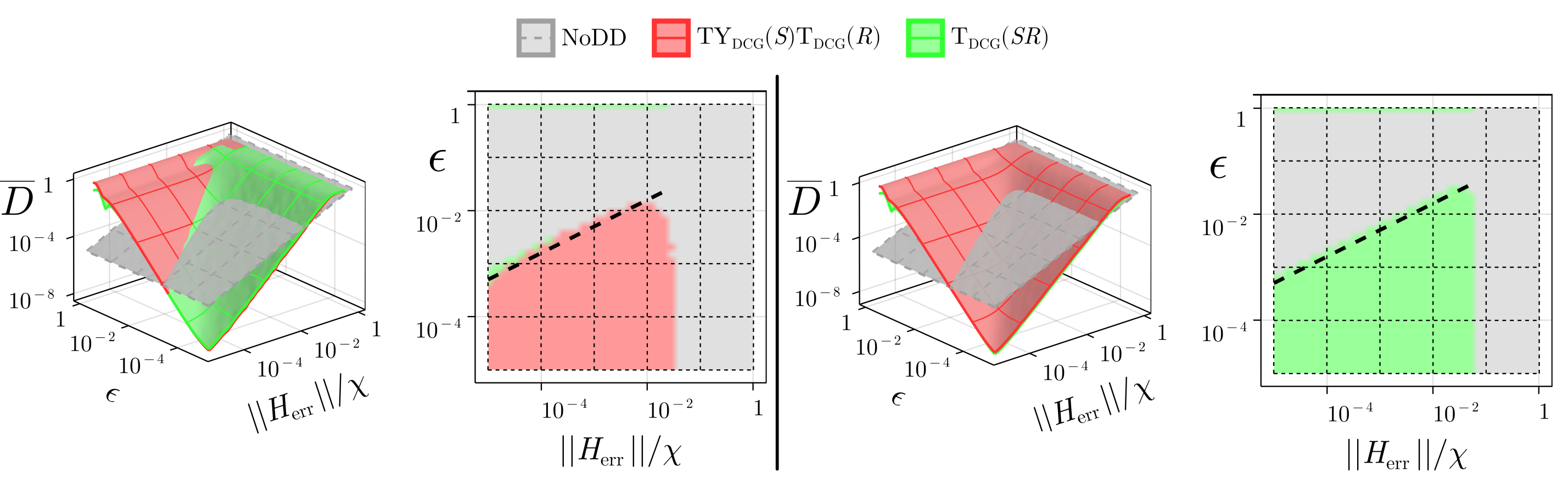}
    \caption{(a) (resp.\ (b)): Average distance in the disorder-dominated (resp.\ interaction-dominated) regime in the $(\norm{H_{\mathrm{err}}}/\chi,\epsilon)$ parameter space, where $\norm{H_{\mathrm{err}}} = \norm{H_{\mathrm{dis}} + H_{\mathrm{dd}}}$ and $\norm{H_{\mathrm{dd}}}/\norm{H_{\mathrm{dis}}} = 10^{-1}$ (resp.\ $\norm{H_{\mathrm{dis}}}/\norm{H_{\mathrm{dd}}} = 10^{-1}$), in the case of flip-angle errors in the DD pulses.}
    \label{fig.DDerrors}
\end{figure}

\par The results are presented in Fig.~\ref{fig.DDerrors}a and Fig.~\ref{fig.DDerrors}b for the disorder-dominated and interaction-dominated regimes, respectively, and show that the DCGs still offer some protection whenever $\epsilon \lesssim 10^{-0.8}\sqrt{\norm{H_{\mathrm{err}}}/\chi}$. When flip-angle errors are greater than this critical value, the DCGs introduce more errors than they correct. Note that similar results can be obtained for any systematic pulse error which belongs to the correctable subspace of $\mathrm{TEDD}$ and $\mathrm{TEDDY}$, such as axis-misspecification errors, where each rotation is implemented around an axis which deviates from the intended one~\cite{Lidar_2023_survey,read_2024}. The inclusion of small DD pulse errors does not appear to change the overall hierarchy of the protocols: $\mathrm{T_{DCG}(SR)}$ continues to perform best in the interaction-dominated regime, while $\mathrm{TY_{DCG}(S)}\mathrm{T_{DCG}(R)}$ remains the most effective in the disorder-dominated regime. However, for strong pulse errors—specifically in the regime $\norm{H_{\mathrm{err}}}/\chi\lesssim \epsilon \lesssim 10^{-0.8}\sqrt{\norm{H_{\mathrm{err}}}/\chi}$, the green and red surfaces nearly overlap, indicating that the advantage of the individual-pulse protection scheme over the cycle protection becomes less pronounced. This behavior is expected, as the individual-pulse scheme involves a larger number of DD pulses and thus accumulates more errors. Consequently, $\mathrm{T_{DCG}(SR)}$ may become preferable for certain state preparation protocols within this parameter range.

\subsection{Errors in the balanced pair}\label{Sec.DDerrors}

Let us now consider the same control protocol where flip-angle errors also appear in the balanced pair and the NoDD protocol, such that the faulty control profiles of the NoDD pulse, stretched pulse and identity pulse are given by 
\begin{equation}\begin{aligned}
    &\text{NoDD profile}:\quad f_{\mathrm{NoDD}}(t) = (1+\epsilon)f(t), \quad t \in [0,\tau] \\[5pt]
    & \text{Stretched pulse profile}: \quad f_{\mathrm{str}}(t) = \frac{1}{2}(1+\epsilon^{\mathrm{str}})f(t/2), \quad t \in [0,2\tau] \\[5pt]
    & \text{Identity pulse profile}: \quad f_{\mathrm{id}}(t) =\begin{cases}
        (1+\epsilon)f(t) \quad &t \in [0,\tau] \\
        -(1+\epsilon^{\mathrm{id}})f(2\tau - t) \quad &t \in [\tau, 2\tau]
    \end{cases}
\end{aligned}\end{equation}
where the error parameters may differ when the pulse is stretched or when the sign of the Hamiltonian is switched. One can show that, in the case where $\epsilon = \epsilon^{\mathrm{str}}=\epsilon^{\mathrm{id}}$, the balanced pair implements the over-rotation with no additional errors compared to the NoDD case, but do not correct it, such that the DCG's performance is upper-bounded by the flip-angle error but it should still provide some protection whenever the errors of the DD pulses are not significant ($\epsilon\lesssim 10^{-1}\sqrt{\norm{H_{\mathrm{err}}}/\chi}$). 

\par
\begin{figure}[ht]
    \centering
    \includegraphics[width=\linewidth]{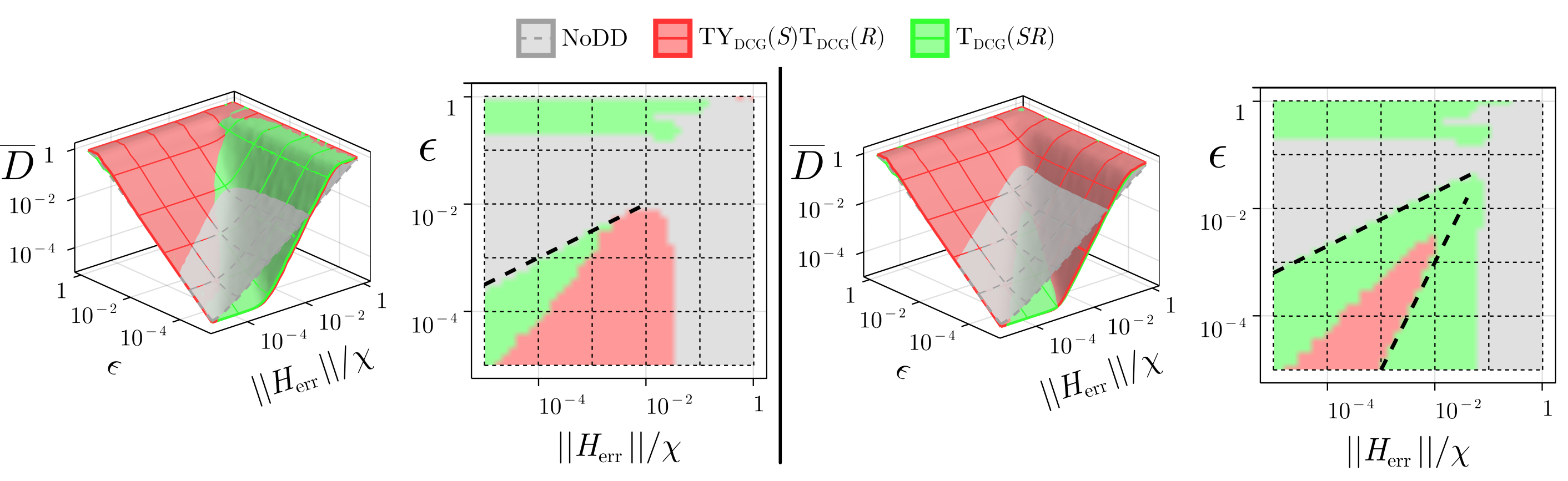}
    \caption{(a) (resp. (b)) : Average distance in the disorder-dominated (resp. interaction-dominated) regime in the $(\norm{H_{\mathrm{err}}}/\chi,\epsilon)$ parameter space, where $\norm{H_{\mathrm{err}}} = \norm{H_{\mathrm{dis}} + H_{\mathrm{dd}}}$ and $\norm{H_{\mathrm{dd}}}/\norm{H_{\mathrm{dis}}} = 10^{-1}$ (resp. $\norm{H_{\mathrm{dis}}}/\norm{H_{\mathrm{dd}}} = 10^{-1}$), in the case of flip-angle errors in the DD pulses and the balanced pair, using $\epsilon = \epsilon^{\mathrm{str}}=\epsilon^{\mathrm{id}}$.}
    \label{fig.BalancedPair.error.}
\end{figure}

 Adding such flip-angle errors to the balanced pair and the NoDD protocol for the same quantum system and control protocol as in Sec.~\ref{Sec.DDerrors}, we find that the flip-angle errors limit the DCG's performance whenever $\epsilon \gtrsim 10^{0.7}\qty(\norm{H_{\mathrm{err}}}/\chi)^2$, greatly reducing the benefit of the DCG strategies (see Fig.~\ref{fig.BalancedPair.error.}). In the regime $10^{-1}\sqrt{\norm{H_{\mathrm{err}}}/\chi} \gtrsim \epsilon \gtrsim 10^{0.7}\qty(\norm{H_{\mathrm{err}}}/\chi)^2$, the two protocols exhibit comparable performance. The cycle-protection scheme performs slightly better when $\epsilon \gtrsim\norm{H_{\mathrm{err}}}/\chi$, whereas the single-pulse protection scheme is slightly superior when $\epsilon \lesssim\norm{H_{\mathrm{err}}}/\chi$.

\par
\begin{figure}[ht]
    \centering
    \includegraphics[width=\linewidth]{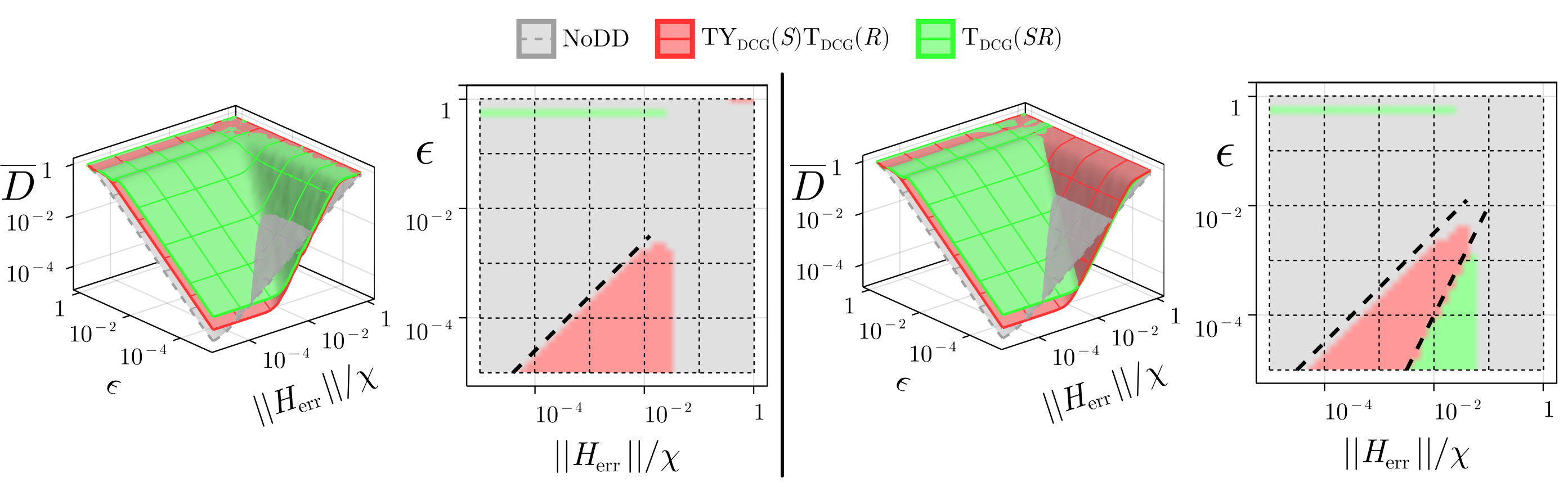}
    \caption{(a) Average distance in the disorder-dominated regime within the $(\norm{H_{\mathrm{err}}}/\chi,\epsilon)$ parameter space, where $\norm{H_{\mathrm{err}}} = \norm{H_{\mathrm{dis}} + H_{\mathrm{dd}}}$ and $\norm{H_{\mathrm{dd}}}/\norm{H_{\mathrm{dis}}} = 10^{-1}$.  
(b) Same as (a), but for the interaction-dominated regime with $\norm{H_{\mathrm{dis}}}/\norm{H_{\mathrm{dd}}} = 10^{-1}$.  
In both cases, flip-angle errors are introduced in the dynamical decoupling pulses and the balanced pair, with $\epsilon = \epsilon^{\mathrm{str}}=-\epsilon^{\mathrm{id}}$.}
    \label{fig.BalancedPair.error.2}
\end{figure}

 In the more general case where $\epsilon\neq\epsilon^{\mathrm{str}}\neq\epsilon^{\mathrm{id}}$, the identity pulses in the balanced pair may introduce additional errors, which are intrinsic to the DCGs and not corrected to first order. This is the case, for instance, when $\epsilon=\epsilon^{\mathrm{str}}=-\epsilon^{\mathrm{id}}$ (see Fig.~\ref{fig.BalancedPair.error.2}), where each identity pulse introduces an additional error into the system. In this case, uncorrected errors appear in the DCG which do not occur in the NoDD protocol, which changes the regimes where NoDD performs better than any DCG strategies to $\epsilon \gtrsim 10^{-0.6}\norm{H_{\mathrm{err}}}/\chi$. In this regime, the DCG strategy which protects rotation and squeezing individually significantly outperforms the other strategy, as this reduces the total time spent applying the error-prone identity pulses which account for the dominant source of pulse errors.

\section{Distance metric and error analysis}\label{Ap.DCg.An}

In the most general case of a system interacting with its environment, a relevant environment-invariant metric which measures the distance between the system+bath evolution and the system-only target evolution is given by~\cite{Grace_2010} 
\begin{equation}
    D(U,V) = \frac{1}{\sqrt{2d_Sd_B}}\min_{\Phi}\norm{U - V\otimes \Phi}_{\mathrm{Fr}} = \sqrt{1 - \frac{1}{d_Sd_B}\norm{\Gamma}_{\mathrm{Tr}}}, 
\end{equation}
where 
\begin{equation}
    \begin{cases}
        &\norm{\Gamma}_{\mathrm{Tr}} = \Tr\qty[\sqrt{\Gamma^{\dagger}\Gamma}]\\[4pt]
        &\Gamma = \tr_S\qty[U(V^{\dagger}\otimes \mathds{1}_B)]
    \end{cases}.
\end{equation}
This metric can be used to evaluate the performance of DD protocols~\cite{Lidar_2013,read_2024}. In the case of a closed system as considered in the main text, the distance metric reduces to 
\begin{equation}
    D(U,V) = \frac{1}{\sqrt{2d_S}}\norm{U - V}_{\mathrm{Fr}} = \sqrt{1 - \frac{1}{d_S}\abs{\Tr\qty[UV^{\dagger}]}}.
\end{equation}
\par Due to finite-duration errors, the propagator $V(t)$ of the pulse $V(\tau)=V$ is replaced by the propagator 
\begin{equation}
    U = Ve^{-i\Phi_V} \quad\mathrm{with}\quad \Phi_V = \sum_{n=1}^{\infty} \Phi_V^{[n]}
\end{equation}
where the finite-duration error operator $\Phi_V$ is expressed as a series by performing a Magnus expansion in the toggling frame with respect to control Hamiltonian which generates $V(t)$. When decoherence is small enough, that is when $\tau \norm{H_{\mathrm{err}}}\ll 1$ where $\norm{H_{\mathrm{err}}}$ is the supremum operator norm of the noise Hamiltonian, finite-duration errors can be approximated by the first term of the series,
\begin{equation}
    \Phi_V \approx \Phi_V^{[1]} = \int_0^{\tau}dt\, V^{\dagger}(t)H_{\mathrm{err}}V(t)
\end{equation}
which corresponds (up to a factor $1/\tau$) to the finite-duration error Hamiltonian given in the main text, see Eq.~\eqref{eq.fin.dur.}. The norm of the higher-order terms of the Magnus expansion can be upper-bounded as~\cite{Blanes_2009}
\begin{equation}
    \norm{\Phi_V^{[n]}} \leq \pi \qty(\frac{\tau\norm{H_{\mathrm{err}}}}{\xi})^{n}
\end{equation}
where $\xi \approx 1.0868$ is a convergence radius. For the second-order term, a tighter upper-bound is given by 
\begin{equation}
    \norm{\Phi_V^{[2]}} \leq \frac{1}{2}\qty(\tau\norm{H_{\mathrm{err}}})^2.
\end{equation}
In this case, one can derive an upper-bound on the distance using a Taylor expansion, 
\begin{equation}\begin{aligned}
    D(U,V) &\lesssim \frac{1}{\sqrt{2d_S}}\norm{V\qty(\mathds{1}_S -i\Phi_V^{[1]} + \dots ) - V}_{\mathrm{Fr}}\\
    & \leq \frac{1}{\sqrt{2}}\norm{\int_0^{\tau}dt\, V^{\dagger}(t)H_{\mathrm{err}}V(t)} \leq \frac{1}{\sqrt{2}} \tau \norm{H_{\mathrm{err}}}
\end{aligned}\label{eq.UBNODD}\end{equation}
where we used the norm inequality $\frac{1}{\sqrt{d_S}}\norm{\bcdot}_{\mathrm{Fr}} \leq\norm{\bcdot} $, the triangle inequality $\norm{A+B}\leq \norm{A}+\norm{B}$, submultiplicativity $\norm{AB}\leq \norm{A}\norm{B}$ and unitary invariance of the supremum operator norm. In the low-decoherence regime, the distance thus scales linearly with the norm of the unwanted noise Hamiltonian.

\subsection{DCG error analysis}

In the case of a DCG, the propagator of the total sequence can still be written as 
\begin{equation}
    U = Ve^{-i\Phi_{\mathrm{DCG}}} \quad\mathrm{with}\quad \Phi_{\mathrm{DCG}} = \sum_{n=1}^{\infty} \Phi_{\mathrm{DCG}}^{[n]}
\end{equation}
where $\Phi_{\mathrm{DCG}}$ is again obtained by performing a Magnus expansion in the toggling frame with respect to the dynamical decoupling sequence~\cite{DCG_2009_PRA,DCG_2009_PRL}. When decoherence is small enough, the series converges and can be approximated by its first term, which reads 
\begin{equation}
     \Phi_{\mathrm{DCG}}^{[1]} = \sum_{\lambda=a,b} \Pi_{\mathcal{G}}\qty(\Phi_{\lambda}) +  \Pi_{\mathcal{G}}\qty(\Phi_{V}) 
\end{equation}
where $\Phi_{\lambda=a,b}$ are the finite-duration errors of the pulses $a$ and $b$ of the DD sequence, $\Phi_V$ is the finite-duration error of the balanced pair and $\Pi_{\mathcal{G}}$ is the DD symmetrization
\begin{equation}
    \Pi_{\mathcal{G}}(S) = \frac{1}{\abs{\mathcal{G}}}\sum_{g\in \mathcal{G}}g^{\dagger}Sg 
\end{equation}
which projects the operator $S$ on a $\mathcal{G}$-invariant subspace of the space of operators, suppressing $S$ entirely in the case where $S$ belongs to the correctable subspace of the group $\mathcal{G}$. In the case where the DD symmetrization suppresses the finite-duration error of the pulses and the balanced pair, $\Phi_{\mathrm{DCG}}^{[1]} = 0$ and if decoherence is small enough, that is $\tau_{\mathrm{DCG}}\norm{H_{\mathrm{err}}}\ll 1$ where $\tau_{\mathrm{DCG}}$ is the duration of the DCG, the finite-duration error is approximated by the second-order term in the expansion. We can then find an upper-bound as 
\begin{equation}\begin{aligned}
    D(U,V) \lesssim \frac{1}{\sqrt{2}}\norm{\Phi_{\mathrm{DCG}}^{[2]}} \leq \frac{1}{\sqrt{2}} \frac{1}{2}\qty(\tau_{\mathrm{DCG}}\norm{H_{\mathrm{err}}})^2.
\end{aligned}\label{eq.UBDCG}\end{equation}
In order to estimate the regime of parameter where the DCG provides an improvement over the NoDD case, one can compare the upper-bounds~\eqref{eq.UBDCG} and~\eqref{eq.UBNODD} and find when the upper-bounds of the DCG is smaller than that of the unprotected pulse, \ie, when 
\begin{equation}
    \tau \norm{H_{\mathrm{err}}} \gtrsim \frac{1}{2} \tau_{\mathrm{DCG}}^2\norm{H_{\mathrm{err}}}^2.
\end{equation}
By defining a parameter $\alpha > 1$ such that $\tau_{\mathrm{DCG}} = \alpha \tau$, this condition is satisfied when
\begin{equation}
    \frac{2}{\alpha^2} > \tau \norm{H_{\mathrm{err}}}.
\end{equation}
As the duration $\tau$ is lower-bounded by the pulse amplitude $\chi$, one can also define some parameter $\gamma$ as $\tau = \gamma/\chi$, such that 
\begin{equation}
    \frac{2}{\gamma \alpha^2} > \frac{\norm{H_{\mathrm{err}}}}{\chi}.\label{eq.convMagnus}
\end{equation}
The regime of parameters where a DCG outperforms NoDD thus depends on the parameter $\alpha$, which quantifies the time-overhead of the DCG. Note that this estimation is not very tight~; for instance, for the $\mathrm{TEDD}$ sequence used to protect a $\gamma=\pi/2$ rotation, the total duration of the DCG is given by 
\begin{equation}
    \tau_{\mathrm{DCG}} = 24\frac{2\pi}{3\chi} + 12\frac{\pi}{\chi}
\end{equation}
where $\frac{2\pi}{3\chi}$ is the time it takes to perform a $2\pi/3$ rotation with a pulse amplitude $\chi$ and $\pi/\chi$ is the time it takes to perform a $\pi/2$ rotation when the pulse is stretched by a factor of two. Dividing $\tau_{\mathrm{DCG}}$ by $\tau = \pi/2\chi$, we find that $\alpha = 76$ and $\norm{H_{\mathrm{err}}}/\chi \lesssim 10^{-4}$, while the DCG is observed to provide an improvement over NoDD in the regime $\norm{H_{\mathrm{err}}}/\chi \lesssim 10^{-1.6}$ in our numerical simulations. A slightly tighter estimate can be found by considering also the second-order term of the Magnus expansion in the NoDD case, but the analytical estimate still differs with the numerical results as the upper-bound on the second-order term $\mathrm{\Phi_{DCG}^{[2]}}$ is itself not very tight and the norm of high-order terms tend to be over-estimated.

\subsection{DCG error analysis with pulse errors}

In the case of control errors in the DD pulses, the Eulerian design of the sequence auto-corrects the errors, such that their introduction only impacts the second order term which now reads 
\begin{equation}
    \Phi_{\mathrm{DCG}}^{[2]} \leq \frac{1}{2}\tau_{\mathrm{DCG}}^2\qty(\norm{H_{\mathrm{err}}} + \chi\epsilon)^2\label{Type1}
\end{equation}
where $\epsilon\ll 1$ is the error parameter defined in the main text and $\chi$ is the pulse amplitude. The parameter regime where the DCG provides an advantage over NoDD is then found by solving the inequality 
\begin{equation}
    \tau \norm{H_{\mathrm{err}}} \gtrsim \frac{1}{2} \tau_{\mathrm{DCG}}^2\qty(\norm{H_{\mathrm{err}}} + \chi\epsilon)^2.
\end{equation}
In the case where $\norm{H_{\mathrm{err}}}/\chi$ is small enough~\eqref{eq.convMagnus}, we find that the parameter regime where the DCG still outperforms NoDD is estimated by
\begin{equation}
    \epsilon \lesssim -\frac{\norm{H_{\mathrm{err}}}}{\chi} + \sqrt{\frac{2}{\gamma\alpha^2}}\sqrt{\frac{\norm{H_{\mathrm{err}}}}{\chi}} \approx \sqrt{\frac{2}{\gamma\alpha^2}}\sqrt{\frac{\norm{H_{\mathrm{err}}}}{\chi}}
\end{equation}
and we recover the power law observed in the numerical calculation (see Appendix~\ref{Ap.C.Error}) and reported in the main text, although the estimate is again not very tight. We can also find the leading error of the DCG by determining the leading term in~\eqref{Type1}, and we find that the leading error of the DCG is caused by the error in the DD pules when $\norm{H_{\mathrm{err}}}/\chi \leq \epsilon$.

\par In the case of errors in the balanced pair, the leading error will now appear in the first-order term of the Magnus expansion in both the NoDD and DCG cases. Considering an error of amplitude $\chi\epsilon$, the NoDD distance upper-bound estimate will be given by 
\begin{equation}
    D(U,V) \lesssim\frac{1}{\sqrt{2}}\tau\qty(\norm{H_{\mathrm{err}}} + \chi\epsilon).
\end{equation}
For the DCG, we should take into account that the uncorrected error only occurs during the balanced pair and not throughout the entire sequence, such that we have 
\begin{equation}
    D(U,V) \lesssim\frac{1}{\sqrt{2}}\qty[\tau_{\mathrm{BP}}\chi\epsilon + \frac{1}{2}\tau_{\mathrm{DCG}}^2\qty(\norm{H_{\mathrm{err}}} + \chi\epsilon)^2 ]\label{type2}
\end{equation}
where $\tau_{\mathrm{BP}}$ is the total duration of all identity pulses and the stretched pulse. Defining $\tau_{\mathrm{BP}}=\beta \tau$ with $\beta>1$, one finds that the DCG outperforms NoDD when 
\begin{equation}
    \epsilon \lesssim \frac{\norm{H_{\mathrm{err}}}}{\chi} - \frac{\beta-1}{\gamma\alpha^2}\qty(-1 + \sqrt{1 + \frac{2\gamma \alpha^2\beta}{(\beta-1)^2}\frac{\norm{H_{\mathrm{err}}}}{\chi}}).
\end{equation}
In the case where decoherence is small enough~\eqref{eq.convMagnus},  the term in parentheses cancels out an we find that the DCG now outperforms NoDD if 
\begin{equation}
    \epsilon \lesssim \frac{\norm{H_{\mathrm{err}}}}{\chi}.
\end{equation}
Note that in the case where the identity pulses add no error to the DCG, we can simply use $T_{\mathrm{BP}}=\tau$ so that $\beta = 1$ and we retrieve the regime 
\begin{equation}
     \epsilon \lesssim \frac{\norm{H_{\mathrm{err}}}}{\chi} - \sqrt{\frac{2}{\gamma\alpha^2}}\sqrt{\frac{\norm{H_{\mathrm{err}}}}{\chi}} \approx  \sqrt{\frac{2}{\gamma\alpha^2}}\sqrt{\frac{\norm{H_{\mathrm{err}}}}{\chi}}.
\end{equation}
We also find that the leading error in the DCG is caused by the control errors whenever 
\begin{equation}
    \epsilon \gtrsim \frac{\gamma\alpha^2}{2\beta}\qty(\frac{\norm{H_{\mathrm{err}}}}{{\chi}}).
\end{equation}

\subsection{Optimal DD pulse amplitudes}

To improve the performance of a DCG, one may increase the amplitude of the DD pulses, thereby shortening the total protocol duration at the cost of introducing larger pulse errors. The optimal pulse amplitude can be estimated by analyzing the upper bound of the protocol’s performance error, assuming systematic pulse errors that are corrected to first order: 
\begin{equation}
    D(U,V) \lesssim \frac{1}{\sqrt{2}}T_{\mathrm{tot}}^2\qty[\norm{H_{\mathrm{err}}} + \chi_{\mathrm{DD}}\eta \epsilon]^2
\end{equation}
where $\chi_{\mathrm{DD}}$ is the DD pulses amplitude, $\epsilon$ is the error parameter, and $\eta \equiv T_{\mathrm{DD}}/T_{\mathrm{tot}}$ the fraction of time spent applying DD pulses, accounting for the fact that errors occur only during those pulses. The total duration of the protocol can be expressed as
\begin{equation}
    T_{\mathrm{tot}} \equiv \frac{\theta_{\mathrm{DD}}}{\chi_{\mathrm{DD}}} + \frac{\theta_{\mathrm{BP}}}{\chi}
\end{equation}
where $\chi$ is the amplitude of the balanced-pair pulses, and $\theta_{\mathrm{DD}}$ ($\theta_{\mathrm{BP}}$) denotes the total rotation (rotation and squeezing) angle implemented by the DD pulses (balanced pairs). Substituting this expression into the upper bound yields 
\begin{equation}
    D(U,V) \lesssim \frac{1}{\sqrt{2}}\theta_{\mathrm{DD}}^2\qty[\frac{\theta_{\mathrm{BP}}}{\theta_{\mathrm{DD}}}\frac{\norm{H_{\mathrm{err}}}}{\chi} + \epsilon + \frac{\norm{H_{\mathrm{err}}}}{\chi_{\mathrm{DD}}}]^2
\end{equation}
The optimal value of $\chi_{\mathrm{DD}}$ minimizes the term in brackets, which can be written as
\begin{equation}
    f(R) = \tilde{h}\left(\frac{1}{R}+\tilde{\theta}\right) + \epsilon
\end{equation}
where $\tilde{h} \equiv \norm{H_{\mathrm{err}}}/\chi$, $R \equiv \chi_{\mathrm{DD}}/\chi\geq 1$ and $\tilde{\theta} \equiv \theta_{\mathrm{BP}}/\theta_{\mathrm{DD}}$.

\par If $\epsilon$ is independent of $R$, e.g., when the errors arise from magnetic-field inhomogeneities across the sample, the upper-bound is minimized when $\chi_{\mathrm{DD}}\to \infty$. However, the improvement in fidelity becomes negligible once 
\begin{equation}
    \begin{aligned}
        f(R)-f(\infty) &\ll f(\infty) \quad
        \Leftrightarrow \quad R\left(\tilde{\theta} + \frac{\epsilon}{\tilde{h}}\right) &\gg 1 
    \end{aligned}
\end{equation}
that is, when further increasing the DD pulse amplitude yields a gain much smaller than the leading error. 
In particular, whenever $\tilde{\theta} + \epsilon/\tilde{h}\gg 1$, \ie when the control errors are too large or $\theta_{\mathrm{BP}}\gg \theta_{\mathrm{DD}}$, increasing $\chi_{\mathrm{DD}}$ offers no significant advantage.

\end{appendix}

\bibliography{refs.bib}

\end{document}